\documentclass[english,11pt]{article}
\usepackage{epsf,amsmath,amssymb,graphicx,dcolumn}
\usepackage{scalefnt}
\usepackage{filemod}
\usepackage{subfigure,ulem}
\usepackage{pstricks}
\usepackage{cite,hyperref}
\usepackage{color}
\usepackage{rotating}
\newcommand{\abbrev}{\rm\scalefont{.9}}
\newcommand{\lhc}{{\abbrev LHC}}
\newcommand{\pt}{\ensuremath{p_T}}
\newcommand{\qt}{\ensuremath{p_T}}

\newcommand{\nll}{{\abbrev NLL}}
\newcommand{\nnll}{{\abbrev NNLL}}
\newcommand{\lo}{{\abbrev LO}}
\newcommand{\nlo}{{\abbrev NLO}}
\newcommand{\nnlo}{{\abbrev NNLO}}
\newcommand{\sm}{{\abbrev SM}}
\newcommand{\qcd}{{\abbrev QCD}}
\newcommand{\thdm}{{\abbrev 2HDM}}
\newcommand{\mssm}{{\abbrev MSSM}}
\newcommand{\fs}[1]{#1{\abbrev FS}}
\newcommand{\mb}{m_{\rm b}}
\newcommand{\mhiggs}{M}
\newcommand{\mgen}{M}
\newcommand{\bbh}{b\bar bH}
\newcommand{\order}[1]{{\cal O}(#1)}
\newcommand{\plus}{{\abbrev +}}
\newcommand{\citere}[1]{Ref.\cite{#1}}
\newcommand{\citeres}[1]{Refs.\cite{#1}}
\newcommand{\eqn}[1]{Eq.\,(\ref{#1})}
\newcommand{\neqn}[1]{Eqs.\,(\ref{#1})}
\newcommand{\fig}[1]{Fig.\,\ref{#1}}

\newcommand{\sct}[1]{Section~\ref{#1}}

\newcommand{\dd}{{\rm d}}
\newcommand{\dy}{{\abbrev DY}}
\newcommand{\ccbar}{c\bar c}
\newcommand{\bbbar}{b\bar b}
\newcommand{\als}{\ensuremath{\alpha_s}}
\newcommand{\bob}{\ensuremath{b_0/b}}
\newcommand{\bobsq}{\ensuremath{b^2_0/b^2}}

\newcommand{\bbosq}{\ensuremath{b^2/b^2_0}}
\newcommand{\hardcoef}[2]{H_{#1}^{#2}}
\newcommand{\chardcoef}[2]{{\cal H}_{#1}^{#2}}
\newcommand{\hardh}[1]{\hardcoef{b}{H#1}}
\newcommand{\chardh}[2]{\chardcoef{\bbbar\leftarrow ij}{\bbh#1}}
\newcommand{\api}{\frac{\alpha_s}{\pi}}
\newcommand{\muF}{\mu_{\rm F}}
\newcommand{\muR}{\mu_{\rm R}}
\newcommand{\Qres}{Q}
\newcommand{\cw}{\ensuremath{\mathcal W}}

\newcommand{\msbar}{\overline{\text{MS}}}

\newcommand{\pdf}{{\abbrev PDF}}
\newcommand{\mstw}{{\abbrev MSTW2008}}
\newcommand{\nnpdf}{{\abbrev NNPDF2.3}}
\newcommand{\cteq}{{\abbrev CT10}}
\newcommand{\Li}{\ensuremath{\text{Li}}}
\newcommand{\bld}[1]{\boldmath{$#1$}}
\newcommand{\lnR}{\ln(\mgen^2/\muR^2)}
\newcommand{\lnthreeQ}{\ln^3(\mgen^2/\Qres^2)}
\newcommand{\lntwoQ}{\ln^2(\mgen^2/\Qres^2)}
\newcommand{\lnQ}{\ln(\mgen^2/\Qres^2)}
\newcommand{\lntwoF}{\ln^2(\mgen^2/\muF^2)}
\newcommand{\lnF}{\ln(\mgen^2/\muF^2)}
\newcommand{\lnQF}{\ln(\Qres^2/\muF^2)}
\newcommand{\lntwoR}{\ln^2(M^2/\muR^2)}

\newcounter{notecount}
\title{\vspace*{-6em}
  \begin{flushright}
    {\small 
      WUB/14-02\\
      ZU-TH 12/14\\
      LPN14-057 \\
    }
  \end{flushright}
\vspace*{2em} {\bf 
Higgs production in bottom quark annihilation: \\ 
Transverse momentum distribution at {\abbrev \bf NNLO+NNLL}}}
\author{ 
Robert V. Harlander$^a$,  Anurag Tripathi$^{b}$, 
Marius Wiesemann$^{c}$\\[2em]
 {\it $^a$Fachbereich C,
  Bergische Universit\"at Wuppertal,}\\[0em] {\it 42097 Wuppertal,
  Germany}\\[1em]
{\it $^b$Dipartimento di Fisica, Universit\`{a} di Torino}\\[0em]
{\it and INFN, Sezione di Torino,}\\[0em]
 {\it 10125 Torino, Italy}\\[1em]
 {\it $^c$Physik-Institut, Universit\"at Z\"urich,}\\[0em]
 {\it 8057 Z\"urich, Switzerland}\\[.3em]
}
\date{}
\begin{document}
\maketitle

\vspace*{1cm}
\begin{abstract}
We present the inclusive transverse momentum distribution for Higgs
bosons produced in bottom quark annihilation at the \lhc{}. The results
are obtained in the five-flavor scheme. The soft and collinear terms at
small $\pt$ are resummed through \nnll{} accuracy and matched to the
\nnlo{} transverse momentum distribution at large $\pt$. We
find that the theoretical uncertainty, derived from a variation of the
unphysical scales entering the calculation, is significantly reduced
with respect to lower orders.
\end{abstract}
\vfill

\thispagestyle{empty}

\clearpage

\section{Introduction}

In the Standard Model (\sm{}), Higgs boson production proceeds
predominantly through gluon fusion. The theoretical efforts that went
into the precise prediction of the corresponding total cross section as
well as kinematical distributions are enormous (see
Refs.\cite{Dittmaier:2011ti,Dittmaier:2012vm,Heinemeyer:2013tqa} for
more information).  Other processes such as associated $V\!H$ or $t\bar
tH$ production, or weak boson fusion, receive their importance from
their characteristic final state particles or kinematics which typically
improve the signal-to-background ratio relative to gluon fusion.

Similar to $t\bar tH$ production, the Higgs boson can also be produced
in association with bottom quarks ($\bbh{}$). Until now, however, this
process has been largely disregarded in \sm{} Higgs searches and
studies, even though its cross section is larger than for $t\bar tH$
production\,\cite{Spira:1997dg}, since the suppression by the smaller
Yukawa coupling is overcompensated by the increased phase space.
However, in searches for a \sm{} Higgs boson, the experimental
significance of the associated production with bottom quarks suffers
heavily from the enormous \qcd{} background.

In theories with an extended Higgs sector, such as the Two-Higgs-Doublet
Model (\thdm{}) or the Minimal Supersymmetric \sm{} (\mssm{}), the
bottom Yukawa coupling can be enhanced relative to the \sm{} so that
$\bbh$ can become the dominant Higgs production mechanism.  Concerning
the theoretical prediction for this process, mainly two complementary
approaches have been pursued in the past. In the {\it four-flavor
scheme} (\fs{4}), the leading-order (\lo{}) partonic processes are $q\bar
q\to b\bar bH$ and $gg\to b\bar bH$, where $q\in\{u,d,c,s\}$. This
approach is most suitable when the bottom quarks are considered as part
of the signature.  The theoretical prediction is available through
next-to-\lo{} (\nlo{}) \qcd{} in the
\fs{4}\cite{Dittmaier:2003ej,Dawson:2005vi,Dawson:2003kb}.

In the {\it five-flavor scheme} (\fs{5}) at \lo{}, the final state
bottom quarks are considered as part of the proton remnants, which are
implicitly integrated over in the parton model. The \lo{} process thus
becomes $b\bar b\to H$, which needs to be convolved with appropriate
$b$-quark density functions. The $\bbh$ process evaluated in the \fs{5}
is thus also referred to as {\it bottom quark annihilation}. This
approach is most suitable for the calculation of the $\bbh$ component to
inclusive Higgs production.  Its advantage with respect to the \fs{4} in
this case is that, on the one hand, logarithms of the form
$\ln\mb/\mhiggs$ ($\mb$ is the bottom quark mass, $\mhiggs$ the Higgs
mass) which arise from integrating over the collinear region of the
final state bottom quark momenta, are implicitly resummed through
{\abbrev DGLAP} evolution. On the other hand, due to the much simpler
structure of the \lo{} process, its theoretical prediction can be
obtained at higher perturbative order than for the \fs{4}. Indeed, the
next-to-\nlo{} (\nnlo{}) result for the inclusive total cross section in
the \fs{5} has been known for more than 10
years \cite{Harlander:2003ai}. The theoretical uncertainty, derived from
renormalization and factorization scale variation, is significantly
smaller than in the \fs{4}, in particular, for Higgs masses above
200\,GeV. Experimental analyses are currently based on a pragmatic
combination of the \nlo{} \fs{4} and the \nnlo{} \fs{5} result, as
suggested in \citere{Harlander:2011aa}.

With increasing luminosity, kinematical distributions of the Higgs boson
will become more and more important for the clear identification of this
particle and the search for possible deviations from the \sm{}
predictions. Among the simplest observables in this respect is the
transverse momentum ($p_T$) distribution of the Higgs. Comparison to
theoretical predictions will provide a handle to the precise nature of
the Higgs couplings, for example to gluons
\cite{Harlander:2013oja,Azatov:2013xha,Grojean:2013nya}, where the Higgs-gluon
coupling is mediated through a quark loop. Similarly, the associated
production of a Higgs with bottom quarks plays a central role to measure
the Higgs-bottom Yukawa coupling, in particular in theories where this
coupling is enhanced.

It is well known that fixed-order predictions of the $\pt$ spectrum
break down for small values of $\pt$. A proper theoretical description
in this region can be obtained by a resummation of logarithmic terms in
$\pt$, leading to a reordering of the perturbative series.  At this
point, it is useful to clarify our notation for the perturbative orders
of the $\pt$ distribution. In gluon fusion as well as in $\bbh$ within
the \fs{5}, the kinematics of the \lo{} partonic process is $2\to 1$,
so that the $\pt$ distribution vanishes for $\pt\neq 0$. Quite often one
therefore speaks of the ``\lo{} $\pt$ distribution'' only when an
additional parton is emitted which can balance a finite $\pt$ of the
Higgs. In this paper, however, we will consistently associate the term
``\lo{}'' with the $2\to 1$ process, so that in our notation, the \lo{}
$\pt$ distribution in gluon fusion and \fs{5}-$\bbh$ is $\sim\delta(\pt)$.

In gluon fusion, the $\pt$ distribution has been studied in great
detail. The \nnlo{} result in the heavy-top limit was presented
long ago\,\cite{deFlorian:1999zd,Glosser:2002gm}. Subleading top-mass
effects were calculated in \citere{Harlander:2012hf}. For the
resummation in the small-$\pt$ region, various approaches have been
pursued.  In \citere{Bozzi:2005wk}, a matching procedure between the
resummed next-to-next-to-leading logarithmic (\nnll{}) terms and the \nnlo{} $p_T$ distribution has been
suggested which, when integrated over all $\pt$, reproduces the total
cross section at \nnlo{}. Its application to the gluon fusion process
was implemented in the program {\tt
  HqT}~\cite{Bozzi:2003jy,Bozzi:2005wk,deFlorian:2011xf}, which
calculates the \nnlo{}\plus\nnll{} $p_T$ spectrum of the Higgs in the
limit of an infinitely heavy top mass. The effects of exact top and
bottom masses on the resummed transverse momentum distribution were
studied at \nlo{}\plus\nll{} in
\citere{Mantler:2012bj,Grazzini:2013mca}.\footnote{A similar study was
  pursued in \citere{Banfi:2013eda}, which additionally discusses the
  mass effects on the $p_T^{\text{veto}}$ cross section, based on the
  techniques presented in \citere{Banfi:2012jm}.}

For $\bbh$, the \nnlo{} $\pt$ spectrum of the Higgs for $\pt>0$ in the
\fs{5} was obtained in \citere{Harlander:2010cz,Ozeren:2010qp}. The jet-
and $\pt$-vetoed rate~\cite{Harlander:2011fx,Harlander:2010cz}, as well
as the fully differential cross section \cite{Buehler:2012cu} are also known
up to \nnlo{}. The special case of $H+b$ production was
  considered earlier in \citere{Campbell:2002zm}. So far, resummation of
  the $\pt$ spectrum of the Higgs produced in bottom annihilation has
  been considered only at \nlo{}\plus\nll{}\,\cite{Belyaev:2005bs}. In
  this paper, we present the first result of the resummed
  \nnlo{}\plus\nnll{} transverse momentum distribution in the \fs{5}.

The remainder of the paper is organized as follows: In
\sct{sec:elements}, we give a brief outline of the $\pt$ resummation
formalism for the production of an uncolored final state. This section
also defines the notation for the rest of the
article. \sct{sec:matching} describes the matching procedure to the
fixed-order result. In \sct{sec:H2}, we present our result for the
so-called {\it hard coefficient} which was the only missing ingredient
for the calculation of the resummed $\pt$ distribution at
\nnlo{}\plus\nnll{}. Our numerical results are presented in
\sct{sec:results}, including a description of the consistency checks
that have been performed on the implementation (\sct{sec:checks}), the
default input parameters (\sct{sec:input}), and finally the
$\pt$ distributions (\sct{sec:NNLL}) for the \lhc{} at a center-of-mass
energy of 8\,TeV (results for 13\,TeV are presented in
\ref{app:results13}). We analyze the dependence of the differential
cross section on the unphysical scales and the parton distributions.
\sct{sec:conclusions} contains our conclusions. In \ref{app:mellins}, we
give complementary information on complex Mellin transforms of some
transcendental functions, that appear in our calculation.

\section{Transverse momentum resummation}


\subsection{Elements}
\label{sec:elements}
For the following discussion, it will be convenient to consider the
production of a general colorless particle of mass $\mgen$ with
transverse momentum $\pt$ in proton-proton collisions. The
specialization to $\bbh$, where $\mgen=M_H$, will be done in
\sct{sec:H2}.

If $\pt$ is significantly smaller than $\mgen$, large logarithms of
$\pt/\mgen$ arise in the distribution $\dd\sigma/\dd\pt$ due to an
incomplete cancellation of soft and collinear contributions.  Since each
order of perturbation theory introduces additional powers of these
logarithms, the na\"ive perturbative expansion in $\als$ is no longer
valid as $\pt\to 0$. However, factorization of soft and collinear
radiation from the hard process allows us to resum the logarithms to all
orders in $\als$.  This factorization is observed when working in the
so-called impact parameter ($b$) space, defined via the Fourier
transformation\footnote{The momentum conservation relates ${\bf p}_T$ to
  the transverse momenta ${\bf K}_T = \sum_i{\bf k}_{i,T}$ of the
  outgoing partons which is factorized in $b$ space using $\delta({\bf
    p}_T + {\bf K}_T) = (2\pi)^{-2}\int\dd{\bf b}\exp[-i{\bf b}\cdot{\bf
      p}_T]\exp[-i{\bf b}\cdot{\bf K}_T]$.}
\begin{equation}
\begin{split}
f({\bf p}_T) = \frac{1}{(2\pi)^2}\int\dd^2 {\bf b}~ e^{-i{\bf b}\cdot {\bf
    p}_T} f({\bf b})\,,
\end{split}
\end{equation}
implying that the limit $\pt \rightarrow 0$ corresponds to $b\to
\infty$.  Using rotational invariance around the beam axis, the angular
integration can be performed, so that we may write the $\pt$
distribution in the form
\begin{equation}
\begin{split}
\label{eq:res}
\frac{\dd \sigma^{F,\text{(res)}}}{\dd \qt^2} = \tau \int_0^{\infty}
\dd b\, \frac{b}{2} \,J_{0}(b \qt) \, W^F(b,\mgen,\tau)\,,
\end{split}
\end{equation}
with the Bessel function $J_0(x)$, $\tau=\mgen^2/S$, and $S$ the
hadronic center-of-mass energy. By the superscript ``(res)'' in
\eqn{eq:res}, we have already indicated that we are going to use this
equation only for $\pt{} \ll M$, where the logarithmically enhanced
terms need to be resummed. The proper inclusion of terms $\pt\gtrsim M$ will be described in
\sct{sec:matching}. Here and in what follows, the superscript $F$ is
attached to process specific quantities; we will set $F=$\,\dy{} for the
Drell-Yan production of a vector boson, for example, and $F=\bbh{}$ for
the $\bbh$ process.

It is convenient to consider the Mellin transform with respect to the
variable $\tau$ of the resummed cross section in $b$ space,
\begin{equation}
\begin{split}
W^F_N(b,\mgen) = \int_0^1\dd\tau \tau^{N-1} W^F(b,\mgen,\tau)\,,
\end{split}
\end{equation}
which can be written
as\cite{Collins:1984kg,Catani:2000vq}\footnote{Throughout this paper,
  parameters that are not crucial for the discussion will be suppressed
  in function arguments.}
\begin{equation}
\begin{split}
 W^F_N(b,\mgen) &= \sum_{c}\hat\sigma^{F,(0)}_{\ccbar}\,
 H_c^F(\als) \\ & \times \exp \left \{- \int_{\bobsq}^{\mgen^2} \frac{\dd
   k^2}{k^2} \Big[ A_c(\als(k)) \ln \frac{\mgen^2}{k^2} + B_c(\als(k)) \Big]
 \right \}\\ & \times \sum_{i,j} C_{ci,N}(\als(\bob)) \, C_{\bar
   {c}j,N}(\als(\bob) ) \, f_{i,N}(\bob) \, f_{j,N}(\bob)\,,
\label{eq:wn}
\end{split}
\end{equation}
where $\hat\sigma_{\ccbar}^{F,(0)}$ is called the Born factor and determines
the parton level cross section at \lo{}.  Unless indicated otherwise,
the renormalization and factorization scales have been set to
$\muF=\muR=\mgen$. The sum $\sum_c$ runs over all relevant quark flavors
$c=q\in\{u,d,s,c,b\}$ and their charge conjugates, as well as gluons,
$c=g$ (where $\bar g\equiv g$). It takes into account that, already at
\lo{}, different initial states can contribute.\footnote{For example, the
  \lo{} \dy{} process receives contributions from all light quark
  flavors.}  In the $\bbh$ process though, only $c\in\{b,\bar b\}$ is
relevant, and
\begin{equation}
\begin{split}
\hat\sigma_{b\bar b}^{\bbh,(0)} = \frac{\pi m_b^2}{6v^2\mhiggs^2}\,,
\label{eq:born}
\end{split}
\end{equation}
where $\mhiggs$ is the Higgs mass, $m_b$ the bottom quark mass, and
$v\approx 246$\,GeV is
the vacuum expectation value of the Higgs field.  The function
$f_{i,N}(q)$ in \eqn{eq:wn} is the Mellin transform of the density
function $f_{i}(x,q)$ of parton $i$ in the proton, where $x$ is the
momentum fraction and $q$ the momentum transfer.  The numerical constant
$b_0=2\exp(-\gamma_E)$, with Euler's constant $\gamma_E= 0.5772\ldots$,
is introduced for convenience.

The perturbative expansion of the {\it resummation coefficients} is
given by
\begin{equation}
\begin{split}
C_{ci,N}(\als) &= \delta_{ci}
+ \sum_{n=1}^{\infty} \left(\api  \right)^n C_{ci,N}^{(n)}\,,\qquad
X(\als) = \sum_{n=1}^{\infty} \left(\api \right)^n
X^{(n)}\,, \\
H_c^F(\als) &= 1+\sum_{n=1}^{\infty} \left(\api \right)^n H_c^{F,(n)}\,,
\label{eq:rescoef}
\end{split}
\end{equation}
where $X\in\{ A_c, B_c\}$. The order at which these coefficients are
taken into account in \eqn{eq:wn} determines the {\it logarithmic
  accuracy} of the resummed cross section; {\it leading logarithmic}
({\abbrev LL}) means that all higher order coefficients except for
$A_c^{(1)}$ are neglected, {\it next-to-{\abbrev\it LL}} (\nll) requires
$A_c^{(2)}$, $B_c^{(1)}$, $C_{ci}^{(1)}$, and $\hardcoef{c}{F,(1)}$,
etc. The coefficients required for the $\bbh$ process at next-to-\nll{}
(\nnll{}) accuracy are given in \sct{sec:H2}.

The fact that the coefficients $A_c$, $B_c$, and $C_{ci}$ in \eqn{eq:wn}
are process independent (i.e., they do not carry a superscript $F$)
assumes a common {\it resummation scheme}\footnote{See
  \citere{Bozzi:2005wk} for details.}  for all ($c\bar c$ initiated,
$c\in\{g,q\}$) processes $F$.  The entire process dependence is then
contained in the {\it hard coefficient} $H_c^F$ and the Born factor
$\hat\sigma_{\ccbar}^{F,(0)}$. In the following, we will work in the
{\it {\abbrev \it DY} scheme}, where $\hardcoef{q}{\dy}\equiv 1$ through
all orders of perturbation theory.  All resummation coefficients are
known in the \dy{} scheme up to the order required in this paper (see
\sct{sec:H2}), with the exception of $H_b^{\bbh}\equiv \hardh{}$ whose
evaluation through \nnlo{} will also be presented in \sct{sec:H2}.

Evolving the parton densities from $\bob$ to $\muF$ in
\eqn{eq:wn} (see \citere{Bozzi:2005wk}), one can define the partonic
resummed cross section $\cw^F_{ij,N}$ through
\begin{align}
W^F_N(b,\mgen) =
\sum_{i,j} 
\cw^F_{ij,N} \left(   b, \mgen, \muF   \right) 
f_{i,N}(\muF) 
f_{j,N}(\muF) \,.
\label{eq:wcurlw}
\end{align}
From a perturbative point of view, $\cw^F$ can
be cast into the form
\begin{equation}
\begin{split}
\cw^F_{ij,N} &\left( b, \mgen, \muF \right)
=\sum_c\hat\sigma^{F,(0)}_{\ccbar}\Bigg\{ {\cal H}^F_{\ccbar\leftarrow
  ij,N} ( \mgen, \Qres, \muF ) + \Sigma^F_{\ccbar\leftarrow ij, N}( L ,
\mgen, \Qres, \muF ) \Bigg\}\,,
\label{eq:curlyW}
\end{split}
\end{equation}
where $L=\ln(Q^2\bbosq)$ denotes the logarithms that are being resummed
in $\cw^F$, and $\Qres$ is an arbitrary {\it resummation scale}.  While
$\cw^F$ is formally independent of $\Qres$, truncation of the
perturbative series will introduce a dependence on this scale which is,
however, of higher order. The variation of the cross section with
$\Qres$ will be taken into account when estimating the theoretical
uncertainty of our final result in \sct{sec:NNLL}. Note that the entire
$b$ dependence, parametrized in terms of $L$, is contained in the functions 
$\Sigma^F_{\ccbar\leftarrow ij}$ which are {\it defined} to vanish at
$L=0$.  Their generic perturbative expansion through \nnlo{}, expressed in
terms of the resummation coefficients of \eqn{eq:rescoef}, can be found
in \citere{Bozzi:2005wk}.  The {\it hard-collinear function}
$\chardcoef{\ccbar\leftarrow ij}{F}$ depends on the coefficients $H_c^F$
and $C_{ci}$ of \eqn{eq:wn}. For $\muF=\muR=Q=\mgen$ and $c\neq g$ (for
$c=g$, see Ref.\,\cite{Catani:2010pd})
\begin{align}
\label{eq:curlyH}
{\cal H}^F_{\ccbar\leftarrow ij,N} = H_c^F(\als)  \,C_{ci,N}(\als)
\,C_{\bar{c}j,N}(\als)\,,
\end{align}
where $\als\equiv \als(\mgen)$.  The expression for
$\chardcoef{\ccbar\leftarrow ij}{F}$ for the $\bbh$ process including
the full scale dependence through \nnlo{} is given in \eqn{eq:cH2}.

Recalling that the formalism discussed in this section is valid only in
the small-$\pt$ region, it is convenient to replace \cite{Bozzi:2005wk}
\begin{equation}
\begin{split}
L\to \tilde L \equiv \ln\left(\frac{Q^2b^2}{b_0^2}+1\right)\,,
\label{eq:Ldef}
\end{split}
\end{equation}
in the resummed cross section, which will prove useful in the next
section to suppress the impact of $\Sigma^F_{\ccbar\leftarrow ij}$ in
the large-$\pt$ region without affecting the logarithmic accuracy under
consideration.  Note, however, that this replacement changes the $\Qres$
dependence of $\Sigma^F_{\ccbar\leftarrow ij}$, so that
\eqn{eq:curlyW}---and therefore
$\dd\sigma^\text{(res)}/\dd\pt^2$---becomes explicitely $\Qres$
dependent. We will come back to this issue in the next section.


\subsection{Matching with the large-\bld{\pt} region}
\label{sec:matching}
In the previous section we recalled the formalism of transverse momentum
resummation at small $\pt$.  In order to obtain a result that is valid
for arbitrary values of $\pt$, a matching to the distribution at high
values of $\pt$ is required, which is predominantly given by the fixed-order 
result. We will follow the additive matching procedure of
\citere{Bozzi:2005wk}, where the resummed-matched result
$[\dd\sigma]_{\text{f.o.}+\text{l.a.}}$ is obtained by subtracting from
the fixed-order distribution $[\dd\sigma]_\text{f.o.}$ the logarithms at
$\pt\to 0$ at the same order in $\als$, and adding the resummed
expression at the appropriate logarithmic accuracy
$[\dd\sigma^\text{(res)}]_\text{l.a.}$:
\begin{align}
\label{eq:match}
\left[\frac{\dd\sigma^F}{\dd\pt^2}\right]_{\text{f.o.}+\text{l.a.}}=
\left[\frac{\dd\sigma^F}{\dd\pt^2}\right]_{\text{f.o.}}
-\left[\frac{\dd\sigma^{F,\text{(res)}}}{\dd\pt^2}\right]_{\text{f.o.}}
+\left[\frac{\dd\sigma^{F,\text{(res)}}}{\dd\pt^2}\right]_{\text{l.a.}}\,.
\end{align}
The logarithmic terms $[\dd\sigma^\text{(res)}]_\text{f.o.}$ are
obtained from the perturbative expansion of \eqn{eq:res}.  The matching
condition is imposed by requiring
\begin{equation}
\begin{split}
\left[ \left[\frac{\dd\sigma^{F,\text{(res)}}}{\dd\pt^2}\right]_{\text{l.a.}}
  \right]_{\text{f.o}} =
\left[\frac{\dd\sigma^{\text{F,(res)}}}{\dd\pt^2}\right]_{\text{f.o.}},
\label{eq:matchcond}
\end{split}
\end{equation}
which defines the logarithmic accuracy needed at each perturbative order
in $\als$ and vice versa. Thus, at a given order in $\als$, it
determines to which order the resummation coefficients of
\eqn{eq:rescoef} are required. Note that the $\Qres$ dependence of
$\dd\sigma^\text{(res)}$ introduced by the replacement $L\to \tilde L$
of \eqn{eq:Ldef} cancels up to higher orders in \eqn{eq:match}.

Integrating \eqn{eq:res} (with $L\to \tilde L$) over $\pt^2$ by using
$\int \dd\pt^2J_0(b\pt) = \pi\delta(b^2)$ (or elementary
properties of the Fourier transform) and $\Sigma_{\ccbar\leftarrow
  ij}(\tilde L=0)=0$, it directly follows that
\begin{equation}
\begin{split}
  \int \dd\pt^2\,
  \frac{\dd\sigma^{F,\text{(res)}}}{\dd\pt^2}=\tau\sum\limits_{c,i,j}\,
  \hat\sigma_{\ccbar}^{F,(0)}\,(\chardcoef{\ccbar\leftarrow ij}{F}\otimes
  f_{i}\otimes 
  f_{j})(\tau)\,,
\label{eq:intres}
\end{split}
\end{equation}
where the convolution of two functions $h_1$ and $h_2$ is defined as
\begin{equation}
\begin{split}
(h_1\otimes h_2)(\tau) \equiv \int_0^1\dd z_1\int_0^1\dd
  z_2\delta(\tau-z_1z_2) h_1(z_1) h_2(z_2)\,.
\end{split}
\end{equation}
Needless to say, $f_i$ and $\chardcoef{\ccbar\leftarrow ij}{F}$ in
\eqn{eq:intres} denote the inverse Mellin transforms of $f_{i,N}$ and
$\chardcoef{\ccbar\leftarrow ij,N}{F}$.

For the r.h.s.\ of \eqn{eq:intres}, it is $[\,\cdot\,]_\text{f.o} =
[[\,\cdot\,]_\text{l.a.}]_\text{f.o.}$; using \eqn{eq:matchcond}, it is
thus easy to see that the integral over $\pt^2$ is the same for
$[\dd\sigma^\text{(res)}]_\text{f.o.}$ and
$[\dd\sigma^\text{(res)}]_\text{l.a.}$.  One therefore obtains a {\it
  unitarity constraint} on the resummed-matched cross section which
implies that the integral over $\pt^2$ reproduces the total cross
section $\sigma_\text{tot}$ at fixed
order:\footnote{\label{foot:deltapt}Note that this line of argumentation
  assumes that terms propotional to $\delta(\pt^2)$ are included in
  $[\dd\sigma^{F,\text{(res)}}]_\text{f.o.}$ as well as
  $[\dd\sigma^F]_\text{f.o.}$ (implying also that logarithms are actual
  plus-distributions). In the practical application of \eqn{eq:match},
  however, these terms cancel and can be disregarded.}
\begin{align}
\int \dd\pt^2 \left[\frac{\dd\sigma^F}{\dd\pt^2}\right]_{\text{f.o.+l.a.}}
&\equiv\left[\sigma^F_\text{tot}\right]_{\text{f.o.}}.
\label{eq:unitarity}
\end{align}
This relation will be used in \sct{sec:H2} to determine the second-order 
coefficient of the hard function $\hardh{}$ numerically, which is
the only missing piece for carrying out the full \nnll{}
$\pt$ resummation for the $\bbh$ process in the \fs{5}.


\subsection{Resummation coefficients and determination of \bld{\hardh{,(2)}}}
\label{sec:H2}
In the \dy{} scheme, the resummation coefficients relevant for the
$\bbh$ process read
\begin{equation}
\begin{split}
A^{(1)}_b &= C_F\,,\qquad A^{(2)}_b = \frac{1}{2} \,C_F \left[
  \left(\frac{67}{18} -\frac{\pi^2}{6} \right)C_A -\frac{5}{9} N_f
  \right]\,,\\ 
  A^{(3)}_b &=
C_A^2 C_F\left(\frac{11 \pi
   ^4}{720}-\frac{67 \pi ^2}{216}+\frac{245}{96}+\frac{11}{24}\zeta_3\right)+C_A C_F N_f \left(\frac{5 \pi
   ^2}{108}-\frac{209}{432}-\frac{7}{12} \zeta_3\right)\\
   &+C_F^2 N_f \left(-\frac{55}{96}+\frac{1}{2}\zeta_3\right)-\frac{1}{108}C_FN_f^2
   + 8\beta_0\,C_F \left(C_A \left(\frac{101}{216}-\frac{7}{16}\zeta_3\right)-\frac{7}{108} N_f\right)
\\
  B^{(1)}_b &=
-\frac{3}{2}\, C_F\,,\qquad B_{b}^{(2)} = \frac{C_F}{4} \,\Bigg[ C_F
  \left(\pi^2 -\frac{3}{4} -12 \zeta_3 \right) + C_A \left( \frac{11}{9}
  \,\pi^2 - \frac{193}{12} + 6\, \zeta_3\right) \\ & \hspace{4.4cm}+ N_f
  \left(\frac{17}{6} -\frac{2}{9} \pi^2 \right) \Bigg]\,,
\end{split}
\end{equation}
where $\beta_0=(11\,C_A-2\,N_f)/12$, $C_F=4/3$, $C_A=3$, and $N_f=5$ is the number of active quark
flavors; furthermore $\zeta_3\equiv\zeta(3)=1.20206\ldots$ with
Riemann's $\zeta$ function.  $A_c^{(n)}$ and $B_c^{(1)}$ are actually
resummation scheme independent. Through $n\in\{1,2\}$, their expressions
have been known for some time \cite{Kodaira:1981nh,Catani:1988vd}, while
$A^{(3)}$ has recently been calculated in\,\citere{Becher:2010tm}. The
coefficient $B_b^{(2)}$ was first obtained in
\citere{Davies:1984hs}.

The $C$ coefficients which arise in our calculation are of the form
$C^{(n)}_{bi}$ ($n\leq 2$), the index $b$ denotes the bottom quark,
and $i\in\{u,d,s,c,b,g\}$. Of course, the respective coefficients
for the charge conjugate partons are also implied in this notation. In $z$
space (i.e., inverse Mellin space), the first-order coefficients in the
\dy{} scheme read \cite{Davies:1984hs}
\begin{equation}
\begin{split}
C_{bg}^{(1)}(z) &= \frac{1}{2}\, z (1-z)\,,\qquad
 C_{bq}^{(1)}(z) =C_{\bbbar}^{(1)}(z) = 0\,,\\
C_{bb}^{(1)}(z) &= \frac{C_F}{2} \left[ \left( \frac{\pi^2}{2} -4
  \right)\delta(1-z) + 1-z \right]\,.
\end{split}
\end{equation}
The off-diagonal \nlo{} coefficients
$C^{(1)}_{bg},C^{(1)}_{bq},C^{(1)}_{\bbbar}$ ($q\in\{u,d,s,c\}$) are
resummation scheme independent.  The second-order coefficients
$C^{(2)}_{bi}$ can be found in \citere{Catani:2012qa}.

Finally, we need to determine the hard coefficient $\hardh{}$ for the
bottom annihilation process in the \dy{} scheme.  At \nlo{}, it can
easily be deduced from the first-order $C$ coefficient in the \dy{}
scheme \cite{Davies:1984hs} and in the $\bbh{}$ scheme
\cite{Ozeren:2010qp},\footnote{A general result for $C^{(1)}_{ij}$ as a
  function of the finite part of the one-loop corrections has also been
  known for some time \cite{deFlorian:2001zd}.} leading to $\hardh{,(1)}
= 3\,C_F$. On the other hand, we have
calculated the \nnlo{} term $\hardh{,(2)}$ in two independent ways.

\paragraph{Numerical evaluation.}
Using \neqn{eq:match}, \eqref{eq:intres} and \eqref{eq:unitarity}, one
finds
\begin{align}
\label{eq:H2calc}
\tau\sum_{ij}\hat\sigma_{b\bar b}^{\bbh,(0)}\,([\chardh{}{,ij}]_{\text{f.o.}}\otimes
f_{i}\otimes
f_{j})(\tau)=\left[\sigma^{\text{(tot)}}\right]_{\text{f.o.}}-\int
\dd\pt^2\,\left[\frac{\dd\sigma^{\text{(fin)}}}{\dd\pt^2}\right]_{\text{f.o.}}\,,
\end{align}
where
$[\dd\sigma^{(\text{fin})}]_{\text{f.o.}}\equiv[\dd\sigma]_{\text{f.o}}
-[\dd\sigma^{(\text{res})}]_{\text{f.o.}}$.  \eqn{eq:H2calc} holds order
by order in $\als$ and for each channel\footnote{As usual, the
  individual partonic subprocesses (channels) $ij$ are defined
  according to the $\msbar{}$ factorization scheme.}
separately.

If we consider $ij=\bbbar$ at $\order{\als^2}$, the only unknown
in \eqn{eq:H2calc} is the hard coefficient $\hardh{,(2)}$, which appears
as a constant in the hard-collinear function $\chardh{,(2)}{,b\bar b}$
(see \eqn{eq:cH2}).  The full $z$ dependence of the latter is
known from the $C$ functions in the \dy{} scheme. Thus, we can simply
fit $\hardh{,(2)}$ using \eqn{eq:H2calc} without any approximations. The
numerical result we obtain is
\begin{align}
\hardh{,(2)} = 10.47 \pm 0.08,
\label{eq:H2a}
\end{align}
where the relatively big uncertainty is caused by the cancellation of
several digits on the right-hand side of \eqn{eq:H2calc}.

\paragraph{Analytic evaluation.} 
\newcommand{\e}{\epsilon} The evaluation of the hard coefficient
requires the knowledge of the purely virtual amplitude for the process
$\bbh$ which was calculated through \nnlo{} in
\citeres{Harlander:2003ai,Ravindran:2006cg}. We give below the {\abbrev UV}
renormalized\footnote{Both $\als$ and $m_b$ are renormalized in the
  $\msbar$ scheme. In particular, we replace $\als$ in the whole amplitude according to Eq.\,(6) of \citere{Ravindran:2006cg}.}  $b\bar bH$ form factor in $d = 4 -2\e$ dimensions
for $\muR = \mgen$, where here and in what follows, $\mhiggs$ denotes
the Higgs boson mass $M_H$.
\begin{equation}
\begin{split}
F_b^h &= \widetilde{F}_b^h + \hat{F}_b^h\,,
\end{split}
\end{equation}
where
\begin{align}
\hat{F}_b^h 
&= \left(\frac{\alpha_s}{\pi}\right) {C_F} \Big[-\frac{1}{2
   \e^2}-\frac{1}{\e}\Big(\frac{{i\pi}}{2}+\frac{3}{4}\Big)+\frac{\pi ^2}{24}\Bigg]
+ {\Big(\frac{\alpha_s}{\pi}\Big)}^2 \Bigg[
{C_F}^2 \Bigg\{\frac{1}{8
   \e^4}+\frac{1}{\e^3}\Big(\frac{{i\pi}}{4}+\frac{3}{8}\Big)
\nonumber \\
&
+\frac{1}{\e^2} \Big(\frac{3 {i\pi}}{8}-\frac{13 \pi
   ^2}{48}+\frac{17}{32} \Big)
+\frac{1}{\e}
\Big(-\frac{5i \pi ^3 }{24}+\frac{{i\pi}}{2}-\frac{4 \zeta (3)}{3}-\frac{5
   \pi ^2}{32}+\frac{53}{64}\Big)
-\frac{7 {i\pi} \zeta (3)}{6}
\nonumber \\
&
+\frac{11 {i\pi}}{8}
-\frac{3i \pi ^3 }{32}-\frac{7 \zeta (3)}{8}+\frac{83 \pi ^4}{960}-\frac{5 \pi ^2}{12}+\frac{7}{4}\Bigg\}
+{C_AC_F} \Bigg\{\frac{11}{32 \e^3}
+\frac{1}{\e^2}
\Big(\frac{11 {i\pi}}{48}+\frac{\pi
   ^2}{96}+\frac{1}{9}
\Big)
\nonumber \\
&
+\frac{1}{\e}
\Big(
\frac{i \pi ^3 }{48}-\frac{67 {i\pi}}{144}+\frac{13 \zeta
   (3)}{16}-\frac{11 \pi ^2}{192}-\frac{961}{1728}
\Big)
+\frac{11i \pi ^3 }{288}+\frac{77 \zeta
   (3)}{144}-\frac{\pi ^4}{288}
\nonumber \\
&
+\frac{67 \pi ^2}{576}-\frac{607}{648}\Bigg\}
+{N_fC_F}
\Bigg\{-\frac{1}{16 \e^3}
+\frac{1}{\e^2}
\left(
-\frac{{i\pi}}{24}-\frac{1}{36}
\right)
+\frac{1}{\e}
\left(
\frac{5 {i\pi}}{72}+\frac{\pi
   ^2}{96}+\frac{65}{864}
\right)
\nonumber \\ 
&
-\frac{i\pi ^3 }{144}-\frac{7 \zeta (3)}{72}-\frac{5 \pi
   ^2}{288}+\frac{41}{324}\Bigg\}\Bigg]
\end{align}
and
\begin{align}
\begin{split}
&\widetilde{F}_b^h = 1+
\api
C_F
\left(\frac{\pi^2}{4}-\frac{1}{2}\right)
+\left(\api\right)^2 \Bigg[{C_A} {C_F} \left(\frac{37 \zeta_3
   }{72}+\frac{83}{144}+\frac{125 \pi ^2}{432}-\frac{\pi
   ^4}{480}\right)
 \\
&+ {C_F^2} \left(-\frac{15 \zeta_3}{8}+\frac{3}{8}+\frac{\pi
   ^2}{24}+\frac{23 \pi ^4}{1440}\right)
+ {C_F} {N_f} \left(\frac{\zeta_3}{9}+\frac{1}{36}-\frac{5 \pi ^2}{108}\right)
 \\
&+ i\pi \left( 
{C_A} {C_F} \left(\frac{13 \zeta_3}{8}-\frac{121 }{216}-\frac{11 \pi^2}{288}\right)+{C_F^2}
   \left(\frac{\pi^2}{8}-\frac{3   \zeta_3}{2}\right)+\left(\frac{7 
   }{54}+\frac{\pi ^2}{144}\right) {C_F} {N_f} \right) \Bigg]\,,
\end{split}
\end{align}
with $\alpha_s = \alpha_s(\mgen)$. All
singular terms are contained in $\hat{F}_b^h$, while $\widetilde{F}_b^h$
remains independent of singularities. Note, however, that $\hat{F}_b^h$
also contains finite terms. The splitting has been done according to
\citere{Catani:2013tia}.
The hard coefficient in the {\it hard scheme} at
$\muR=\mhiggs$ is then obtained at each order in $\alpha_s$
through \cite{Catani:2013tia}
\begin{equation}
\begin{split}
\hardcoef{b,\text{hard}}{H}(\als) = \left|
\widetilde{F}_b^h(\als) \right|^2\,.
\end{split}
\end{equation}
Using the fact that scheme conversion is process independent, i.e.,
\begin{equation}
\begin{split}
\hardcoef{c,\text{hard}}{F} =
\left(1+\Delta_{\text{hard}}\right)\hardcoef{c}{F}\,,
\end{split}
\end{equation}
with an appropriate perturbative factor
$\Delta_\text{hard}=\order{\als}$, and that
$\hardcoef{q}{\dy}(\als)\equiv 1$, the conversion to the \dy{} scheme is
easily carried out using
\begin{equation}
\begin{split}
\hardh{,(1)} &= \hardcoef{b,\text{hard}}{H,(1)} -
\hardcoef{b,\text{hard}}{\dy,(1)}\,, \\
\hardh{,(2)} &= 
\hardcoef{b,\text{hard}}{H,(2)} -
\hardcoef{b,\text{hard}}{\dy,(2)} 
+
(\hardcoef{b,\text{hard}}{\dy,(1)})^2
-
\hardcoef{b,\text{hard}}{H,(1)}
\hardcoef{b,\text{hard}}{\dy,(1)}\,,
\end{split}
\end{equation}
where $H^{\dy}_{b,\text{hard} }$ is the hard coefficient for the
\dy{} process in the hard scheme which is presented in
\citere{Catani:2013tia}. In this way we find
\begin{equation}
\begin{split}
\hardh{,(2)} &= C_F\bigg[
\left( \frac{321}{64} - \frac{13}{48} \pi^2 \right) C_F +
\left(-\frac{365}{288} + \frac{\pi^2}{12} \right) N_f
\\&\qquad
+ \left(\frac{5269}{576} - \frac{5}{12} \pi^2 - \frac{9}{4}
\zeta_3 \right) C_A\bigg]\,.
\label{eq:H2b}
\end{split}
\end{equation}
This yields a numerical value of
$\hardh{,(2)} = 10.52\ldots$, which is in perfect agreement with
\eqn{eq:H2a}. This serves as an important check of our calculation.


\section{Outline of the calculation and results}\label{sec:results}


\label{sec:outcalc}
We are now ready to consider the resummed transverse momentum
distribution of the Higgs boson produced via bottom quark annihilation
through \nnlo{}\plus\nnll{}.  Exemplary Feynman diagrams that enter our
calculation are shown in \ref{app:diag}.  The \lo{} diagram in
\fig{fig:app1f2}\,(a) determines the Born factor given in
\eqn{eq:born}. The virtual one- and two-loop corrections
(e.g.\ \fig{fig:app1f2} (b) and (c)) govern the hard coefficient
$\hardh{}$ as outlined in Section\,\ref{sec:H2}.  \fig{fig:app1f1} shows
a sample of real and mixed real-virtual diagrams that appear at \nnlo{}
for $\pt>0$.  Note that the various subprocesses enter the calculation
at different orders. The $b\bar{b}$ initial state is the only
subprocess present at \lo{}. At \nlo{} the contribution of the
$bg$ channel also has to be taken into account.\footnote{We account all
 charge conjugated and switched initial states to the same subprocess. Thus,
  the $bg$ channel includes $bg$, $\bar{b}g$, $gb$ and $g\bar{b}$.} The
$gg$-, $bb$-, $bq$- and $q\bar{q}$-initiated subprocesses
($q\in\{u,d,s,c\}$) enter only at \nnlo{}. The only subprocess which is
finite at small transverse momenta and needs no resummation is the
$q\bar{q}$ channel.

The calculation of the resummed-matched distribution of \eqn{eq:match}
requires the differential cross section\footnote{The superscript $F$=$\bbh$
  will be dropped in what follows.} $\dd\sigma$ calculated in various
approximations:
\begin{itemize}
\item The analytic transverse momentum distribution at \nnlo{},
  $[\dd\sigma]_\text{f.o.}$, can be taken from \citere{Ozeren:2010qp} (for $p_T>0$, but see footnote\,\ref{foot:deltapt}).
\item The logarithms at \nnlo{},
  $[\dd\sigma^\text{(res)}]_\text{f.o.}$, are obtained
  from the fixed-order expansion of $\dd\sigma^{\text{(res)}}$ which was
  carried out explicitly in Eqs.~(72) and (73) of \citere{Bozzi:2005wk} (again, only $p_T>0$ terms are taken into account).\footnote{The corresponding coefficients are given in 
Eqs.~(63), (64), (66), (67), (68) and (69) of \citere{Bozzi:2005wk}.}
\item For the calculation of the resummed expression,
  $[\dd\sigma^{\text{(res)}}]_{\text{l.a.}}$, we use a modified version of
  the program {\tt HqT}~\cite{Bozzi:2003jy,Bozzi:2005wk,deFlorian:2011xf}, which
  performs the transverse momentum resummation for gluon-induced Higgs
  production in the heavy-top limit. We extended its capabilities to
  also cover the resummation for quark-induced processes and implemented
  the resummation coefficients of the $b\bar{b}H$ process.
\end{itemize}


\subsection{Checks}
\label{sec:checks}
Before presenting numerical results, we comment on various checks that
we made on our calculation and outline our default input parameters. The
analytic $\pt$ distribution at \nnlo{} \cite{Ozeren:2010qp} has been
checked numerically against the partonic Monte Carlo program for $H+$jet
production at the same order of
\citeres{Harlander:2010cz,Harlander:2011fx}, which in turn has been
validated by various related calculations\footnote{For more details see
  also
  \citere{Wiesemann:2012ij}.}\cite{Campbell:2002zm,Frixione:2002ik,Frederix:2009yq,Frixione:2010ra,Hirschi:2011pa}.

The small-$\pt$ behavior of the distribution needs to agree with the
expansion of $\dd\sigma^{\text{(res)}}$.  We checked that the limit
\begin{align}
\left[\frac{\dd\sigma}{d\pt^2}\right]_{\text{f.o.}}\stackrel{\pt\rightarrow 0}{\rightarrow}\left[\frac{\dd\sigma^{\text{(res)}}}{d\pt^2}\right]_{\text{f.o.}}
\end{align}
holds to better than one per-mille in the interval
$0.001$\,GeV$<\pt<0.1$\,GeV.  We also verified that this limit is
independent of the resummation scale.

Furthermore, we used our implementation of $\dd\sigma^{\text{(res)}}$ to
calculate a large number of sampling points in order to approximate the
integral over $\pt$. According to \eqn{eq:intres}, the result has to
yield the (analytically known) hard-collinear function $\chardh{}{}$,
which we verified up to an accuracy of a few per-mille.\footnote{More
  precisely, to verify \eqn{eq:intres} we used resummation scales
  significantly smaller than the mass of the Higgs to reduce the impact
  of resummation at high transverse momenta, because at very high $p_T$
  ($p_T\gtrsim 300$ GeV) the numerical convergence of our implementation
  of $\dd\sigma^\text{(res)}$ deteriorates.} This is quite remarkable,
considering the fact that the determination of $\dd\sigma^{\text{(res)}}$
includes the numerical transform from $b$ to $\pt$ space and from $N$ 
to $z$ space, as well as a fit of the parton distributions in Mellin
space.

We also checked \eqn{eq:unitarity} for the resummed-matched cross
section up to a numerical accuracy considerably better than one
per-mille, using the analytical result for $\chardh{}{}$ as the integral
of $\dd\sigma^{(\text{res})}$. This was already expected from the
agreement between \eqn{eq:H2a} and the analytical result of \eqn{eq:H2b}
for $\hardh{,(2)}$ mentioned above.

All these checks have been performed for various values of the resummation,
factorization, and renormalization scale, separately at order $\alpha_s$
and $\alpha_s^2$, and for the individual partonic subchannels.

At \nlo{}\plus{}\nll{}, the $\pt$ spectrum of the Higgs in $\bbh$ has
already been studied in \citere{Belyaev:2005bs} within the formalism of  \citeres{Collins:1984kg,Nadolsky:2002jr}. Although their approach---in particular, the matching procedure---differs from ours, the
qualitative behavior of our curves is in fairly good agreement at this
order. In particular, we find the same
properties of the resummed-matched curve at high transverse momenta, which is
nontrivial as will be shown in \sct{sec:NNLL}.


\subsection{Input parameters}
\label{sec:input}

We present results for the \lhc{} at $8$ and $13$\,TeV center-of-mass
energy. Our choice for the central factoriza\-tion and renormalization
scale is $\muF=\muR=\mu_0\equiv \mhiggs$; our default value for the
resummation scale is $\Qres=Q_0\equiv \mhiggs/4$. If not stated
otherwise, all numbers are obtained with the {\abbrev
  MSTW2008}~\cite{Martin:2009iq} \pdf{} set, which implies that the
input value for the strong coupling constant is taken as
$\alpha_s\left(M_Z\right)=0.12018$ at \nlo{}, and
$\alpha_s\left(M_Z\right)=0.11707$ at \nnlo{}. For comparison we also
report results for the \nnpdf{} and \cteq{} \pdf{} sets, with their
corresponding $\alpha_s\left(M_Z\right)$ values. Since we are working in
the \fs{5}, the bottom mass is set to zero throughout the calculation,
except for the bottom-Higgs Yukawa coupling which we insert in the
$\msbar$ scheme at the scale $\muR$, derived from the input value
$\mb(\mb)=4.16$\,GeV.

All numbers are evaluated within the framework of the \sm{}. Through
appropriate rescaling of the bottom Yukawa coupling, they are obviously
also applicable to neutral ($CP$ even and odd) Higgs production
within the \thdm{} and, according to the studies of
\citeres{Dawson:2011pe,Dittmaier:2006cz}, even within the \mssm{}.

Sources of theoretical uncertainty and their impact on the numerical
results will be studied in \sct{sec:NNLL}. As usual, the uncertainty due
to the truncation of the perturbative series with respect to $\alpha_s$
will be estimated from the dependence of the cross section on the
unphysical scales $\muF$ and $\muR$. Similarly, the effect of a finite
logarithmic accuracy will be addressed by a variation of
$\Qres{}$. Finally, we will investigate the uncertainty induced by the
\pdf{}s and the input value of $\alpha_s(M_Z)$.


\subsection{Transverse momentum distribution up to  {\abbrev \bf NNLO+NNLL}} 
\label{sec:NNLL}
In this section we present our results for the transverse momentum
distribution of Higgs bosons produced in bottom quark annihilation. We
study the impact of the newly evaluated terms at \nnlo{}\plus\nnll{} by
comparing them to \nlo{}\plus\nll{}, both in absolute size and
in their theoretical uncertainty.

\begin{figure}
\begin{center}
    \begin{tabular}{cc}
     \hspace{-0.45cm}
     \mbox{\includegraphics[height=.3\textheight]{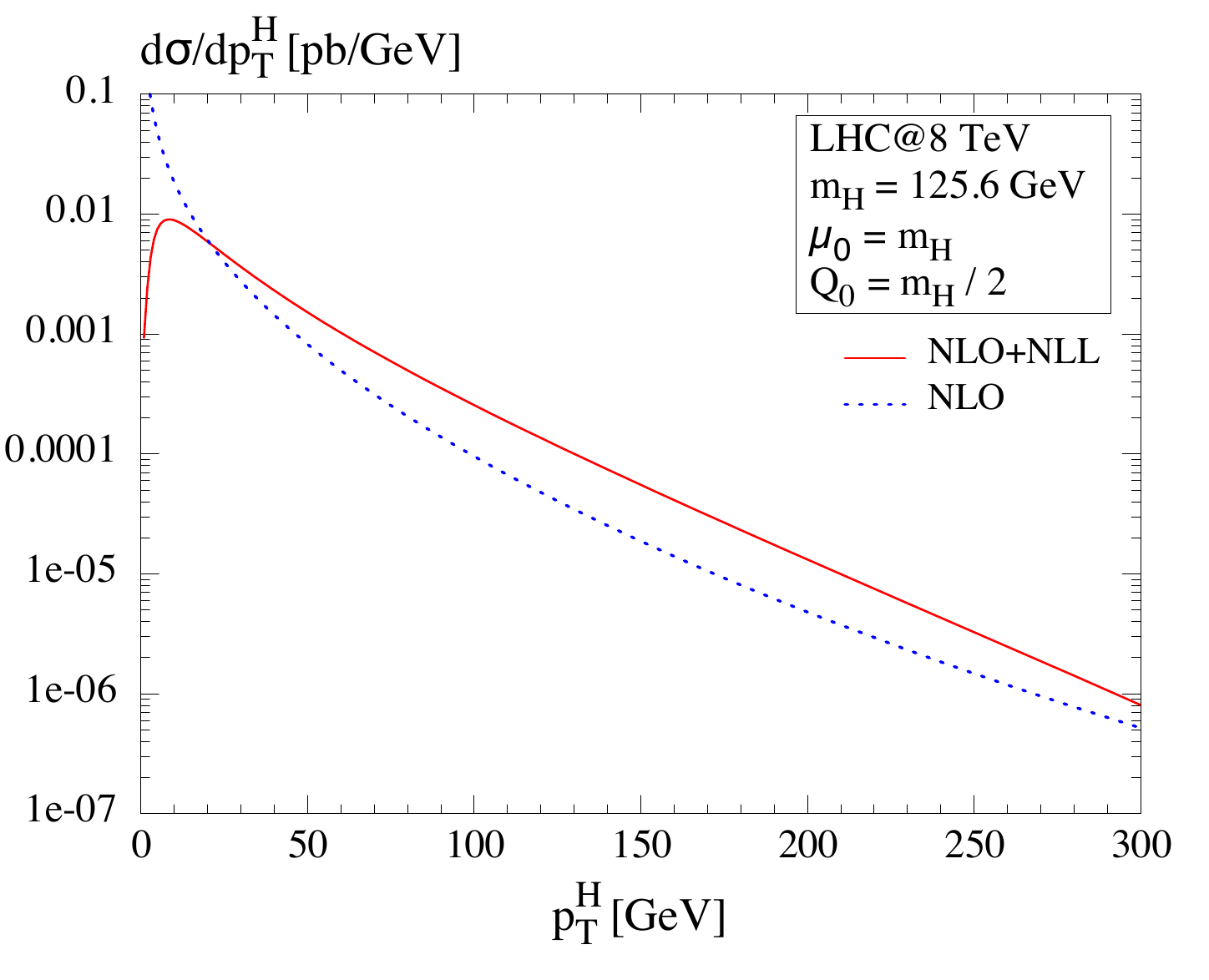}} &
     \mbox{\includegraphics[height=.3\textheight]{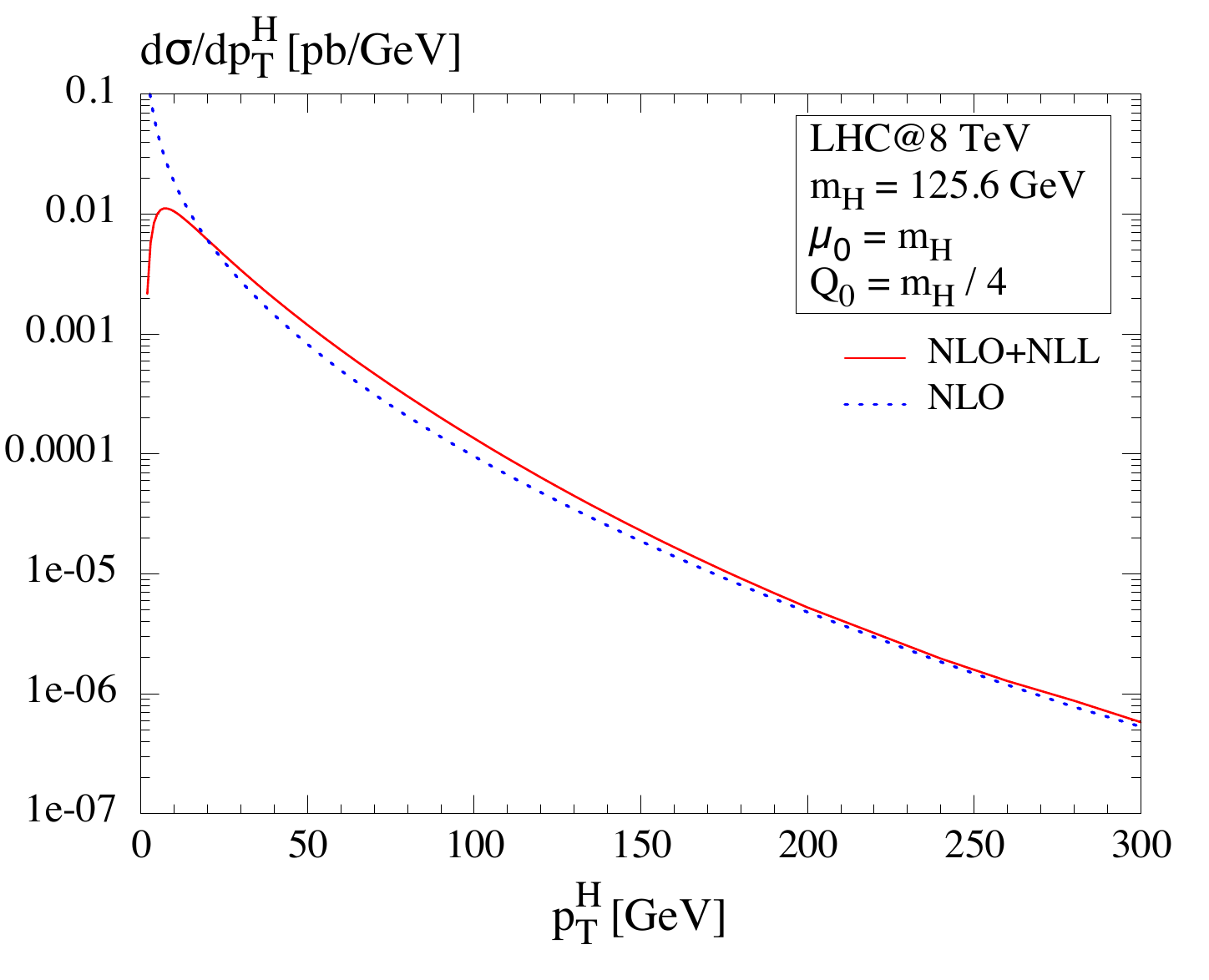}}
            \\
     \hspace{0.15cm} (a) & (b)
    \end{tabular}
    \parbox{.9\textwidth}{%
      \caption[]{\label{fig:nlonll}{Transverse momentum spectrum at
          \nlo{} (blue, dashed line) and at \nlo\plus\nll{} (red, solid line) for
          (a) $Q=M/2$ and (b) $Q=M/4$. (Here and in the following plots,
    $m_H=M$ is the Higgs mass.)}}}
\end{center}
\end{figure}
\fig{fig:nlonll} shows the \nlo{}\plus{}\nll{} together with the fixed-order 
\nlo{} distribution for two values of the resummation scale $Q$:
\fig{fig:nlonll}\,(a) uses $\Qres=M/2$, which is the default value
typically used in gluon fusion\,\cite{Bozzi:2005wk}, while
\fig{fig:nlonll}\,(b) uses $\Qres=M/4$.  Resummation aims at a valid
description of the low-$\pt$ region and indeed, the divergence at
$\pt\to 0$ of the fixed-order result is turned into a regular behavior.
Due to higher-order effects, the fixed-order and the resummed-matched
curve may also significantly differ at $\pt\sim
M$\,\cite{Belyaev:2005bs}, as is also observed for gluon fusion
\cite{Mantler:2012bj,Banfi:2013eda}.\footnote{A standard option in {\tt
    HqT}~\cite{Bozzi:2003jy,Bozzi:2005wk,deFlorian:2011xf}, for example,
  is to use an intersection point between the fixed-order and the
  resummed-matched curve in order to switch from the latter to the
  former towards large $\pt$.} \fig{fig:nlonll} shows that for
bottom quark annihilation, this difference is significantly smaller for
$\Qres=M/4$ than for $Q=M/2$.
This observation motivates us to use $Q=Q_0\equiv M/4$ as the central
resummation scale choice also at \nnlo{}\plus{}\nnll{} in the
following.\footnote{We thank an anonymous referee for this suggestion.}
Nevertheless, for reference, we include results for $Q_0=M/2$ in
\ref{app:Q05}.

\begin{figure}
\begin{center}
    \begin{tabular}{cc}
     \hspace{-0.45cm}
     \mbox{\includegraphics[height=.3\textheight]{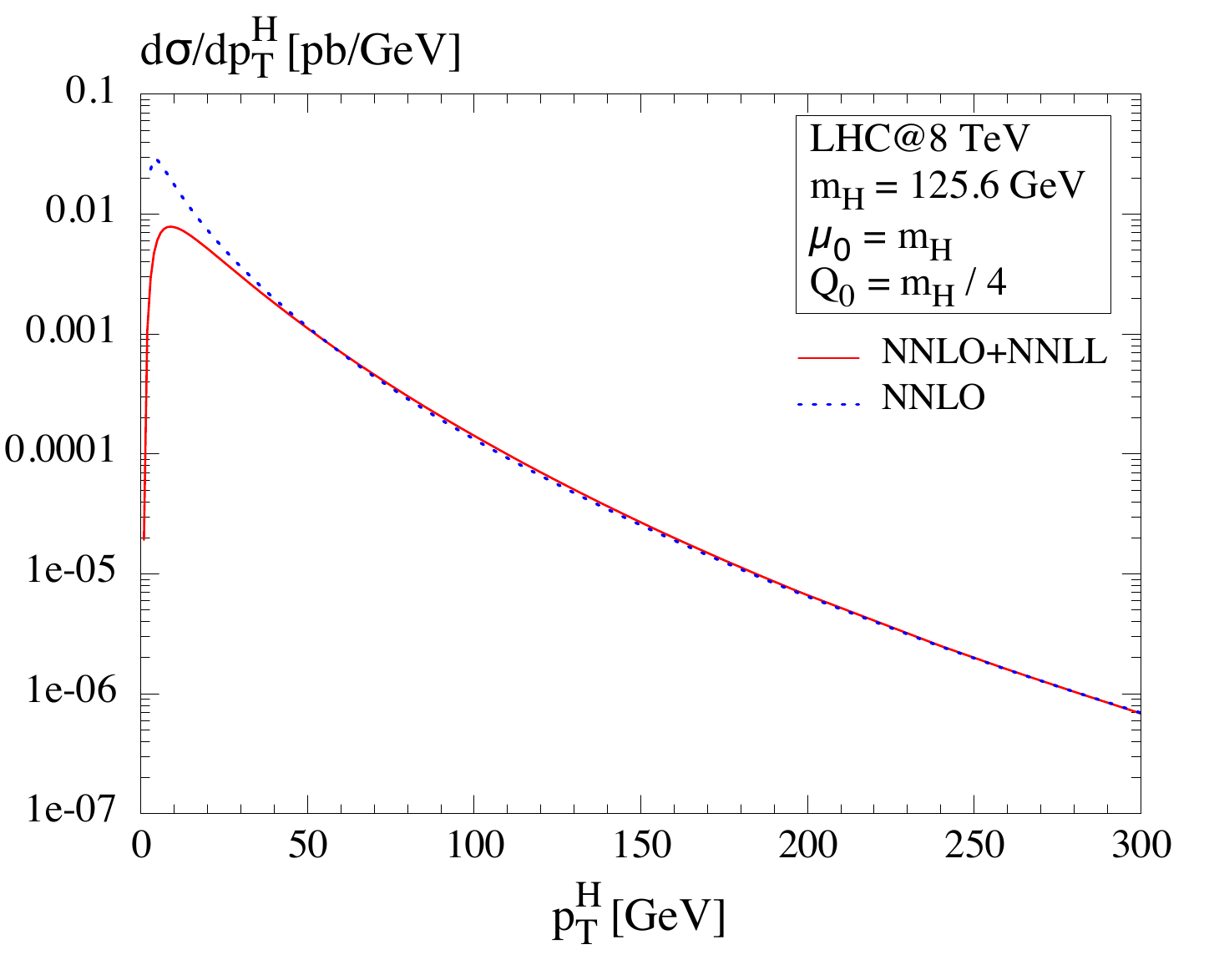}} &
     \mbox{\includegraphics[height=.3\textheight]{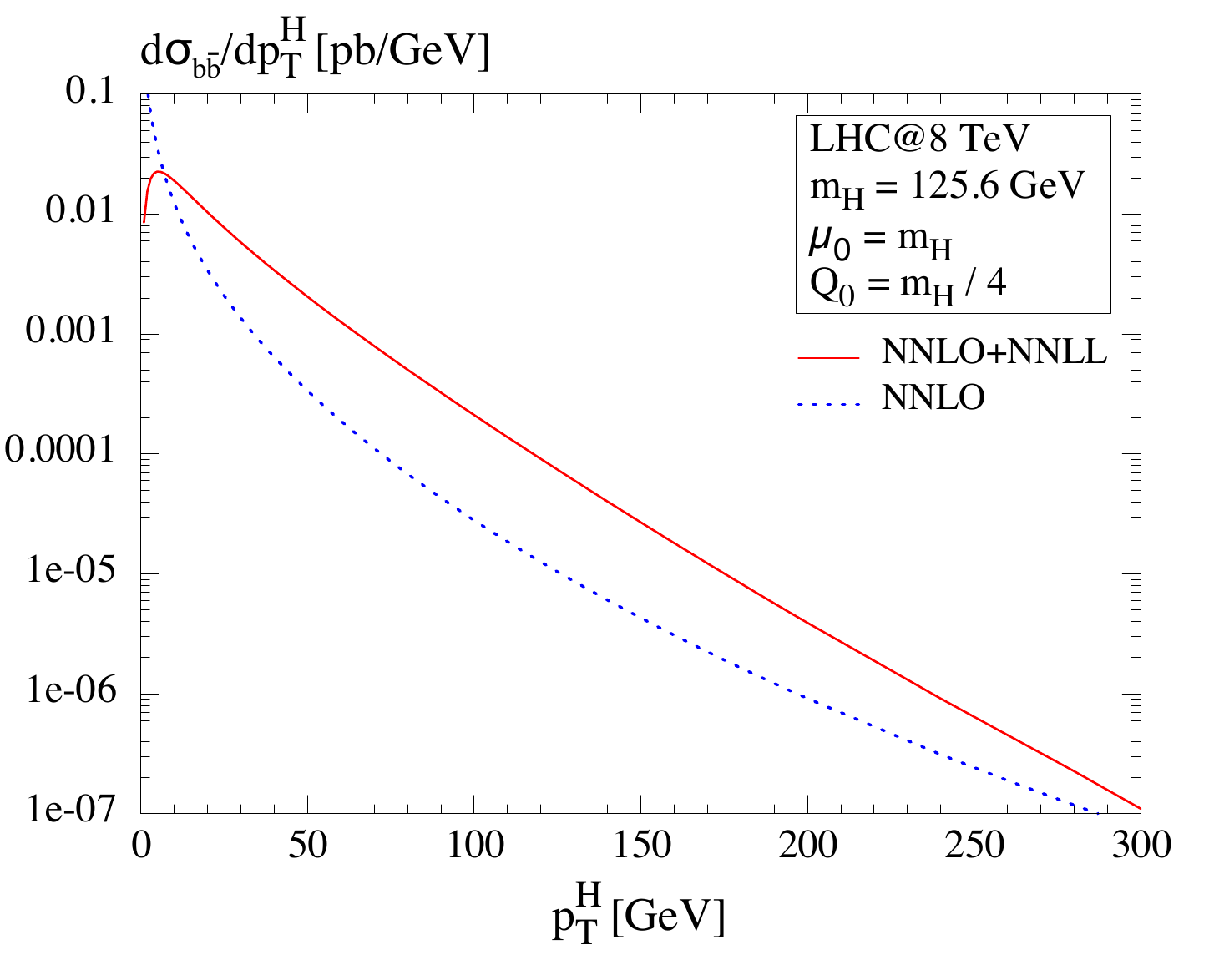}}
            \\
     \hspace{0.15cm} (a) & (b)
    \end{tabular}
    \parbox{.9\textwidth}{%
      \caption[]{\label{fig:highpT}{(a) Transverse momentum spectrum at
          \nnlo{} (blue, dashed line) and at \nnlo\plus\nnll{} (red, solid line)
          for the central scales; (b) only the $b\bar{b}$ channel for
          that quantity. } }}
\end{center}
\end{figure}

\begin{figure}
\begin{center}
\includegraphics[height=.4\textheight]{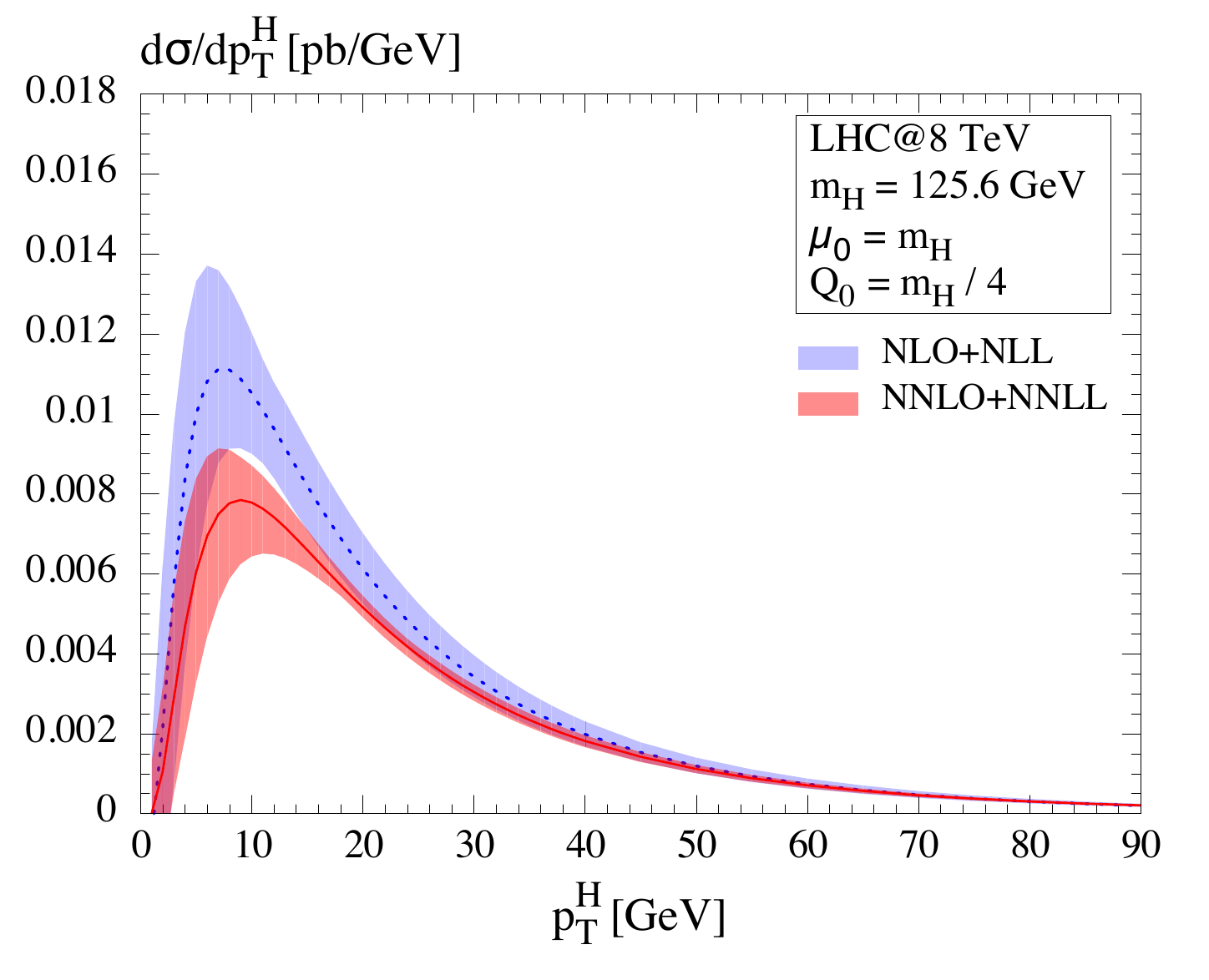}
    \parbox{.9\textwidth}{%
      \caption[]{\label{fig:muFmuR}{Resummed-matched $p_T$ distribution
          at \nlo{}\plus\nll{} (blue, dashed line) and \nnlo{}\plus\nnll{}
          (red, solid line); lines: central scale choices; bands: uncertainty
          due to $\muF{},\muR{}$-variation.  } }}
\end{center}
\end{figure}

At \nnlo{}\plus{}\nnll{}, we find that the agreement between the fixed-order
and the resummed-matched curve is further improved with respect to
\nlo{}\plus\nll{}, see \fig{fig:highpT}\,(a). This confirms that the
difference between these two results at $\pt\sim M$ is due to higher-order 
effects. For $\pt\gtrsim 50$\,GeV, the resummed-matched curve is
practically on top of the fixed-order curve.
We note in passing that the agreement between the fixed-order and the resummed-matched
curve results from nontrivial cancellations among the individual
partonic subchannels. For example, considering only the $b\bar b$
channel, there are still large differences between the two curves, see
\fig{fig:highpT}\,(b), which are, however, compensated by the
   other partonic channels. In conclusion, 
  the \nnlo{}\plus{}\nnll{} result is the first to combine
  the small and high-$p_T$ region in a satisfactory way. This indicates
  its importance to obtain a distribution valid at all transverse
  momenta.

\begin{figure}
\begin{center}
\includegraphics[height=.4\textheight]{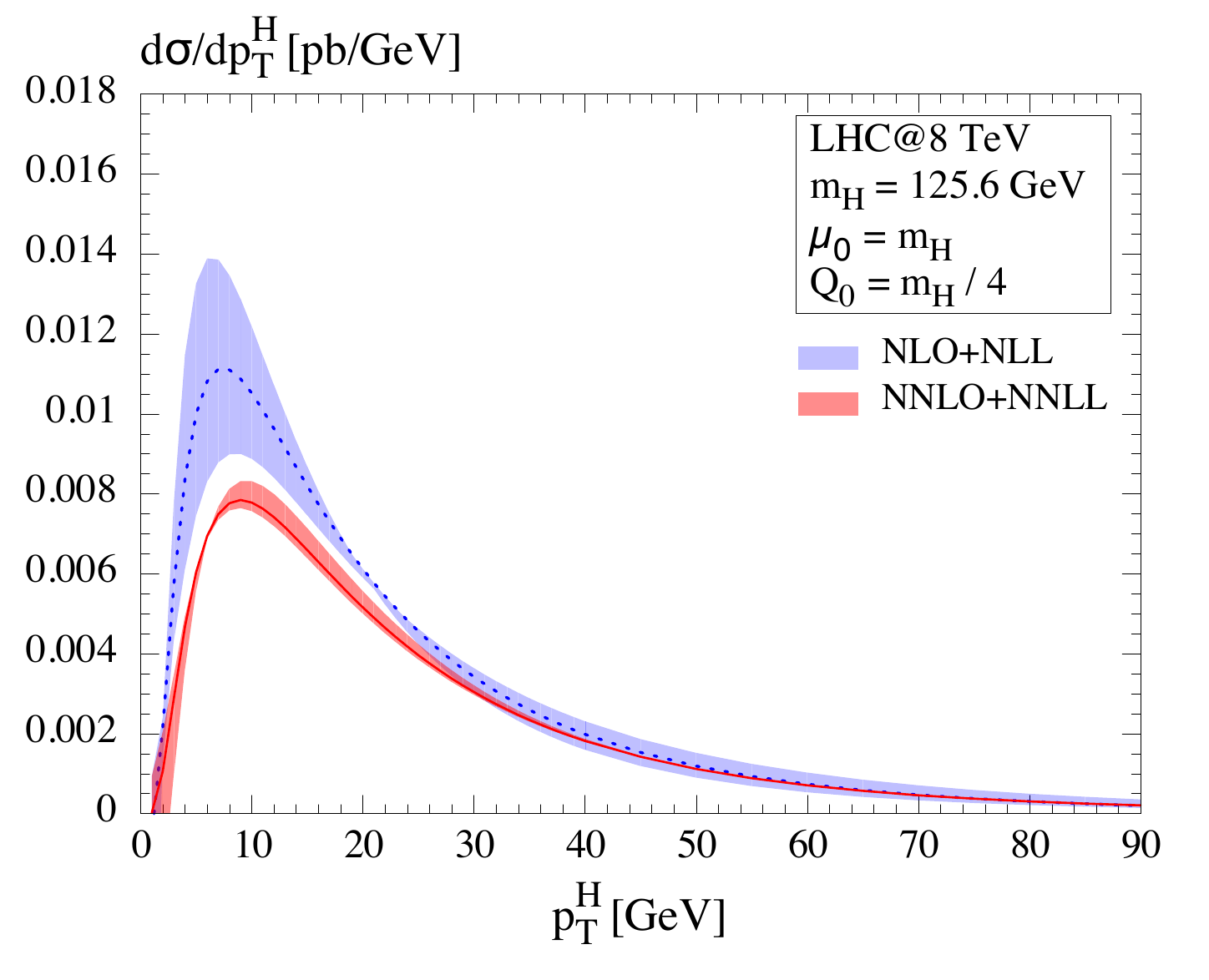}
    \parbox{.9\textwidth}{%
      \caption[]{\label{fig:Qres}{Resummed-matched $p_T$ distribution
          at \nlo{}\plus\nll{} (blue, dashed line) and \nnlo{}\plus\nnll{}
          (red, solid line); lines: central scale choices; bands: uncertainty
          due to $\Qres{}$-variation.  } }}
\end{center}
\end{figure}
\begin{figure}
\begin{center}
\includegraphics[height=.4\textheight]{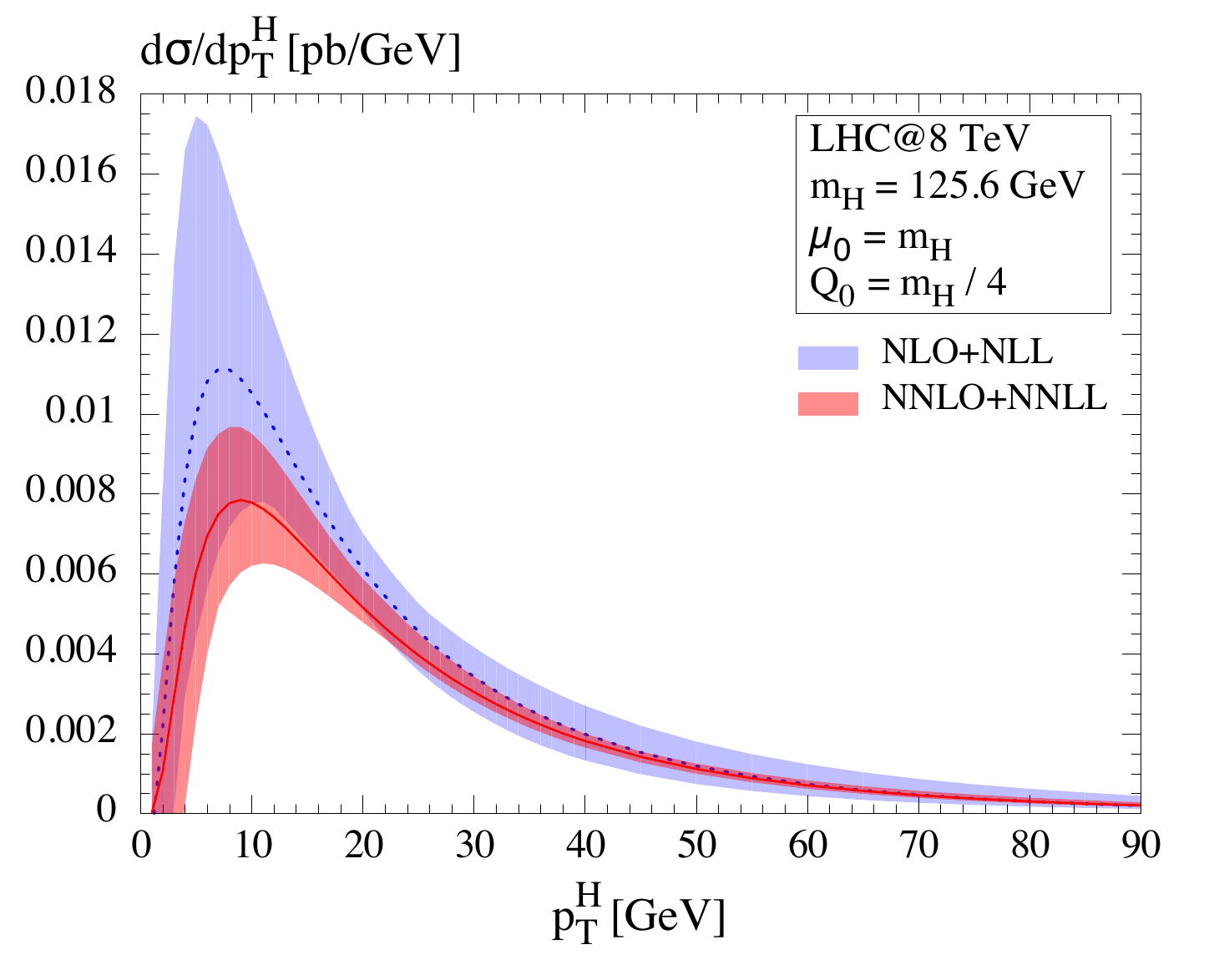}
    \parbox{.9\textwidth}{%
      \caption[]{\label{fig:all}{\sloppy Resummed-matched
          $p_T$ distribution at \nlo{}\plus\nll{} (blue, dashed line) and
          \nnlo{}\plus\nnll{} (red, solid line); lines: central scale
          choices; bands: uncertainty due to variation of all scales.  }
    }}
\end{center}
\end{figure}

\begin{figure}
\begin{center}
    \begin{tabular}{cc}
     \hspace{-0.45cm}
     \mbox{\includegraphics[height=.3\textheight]{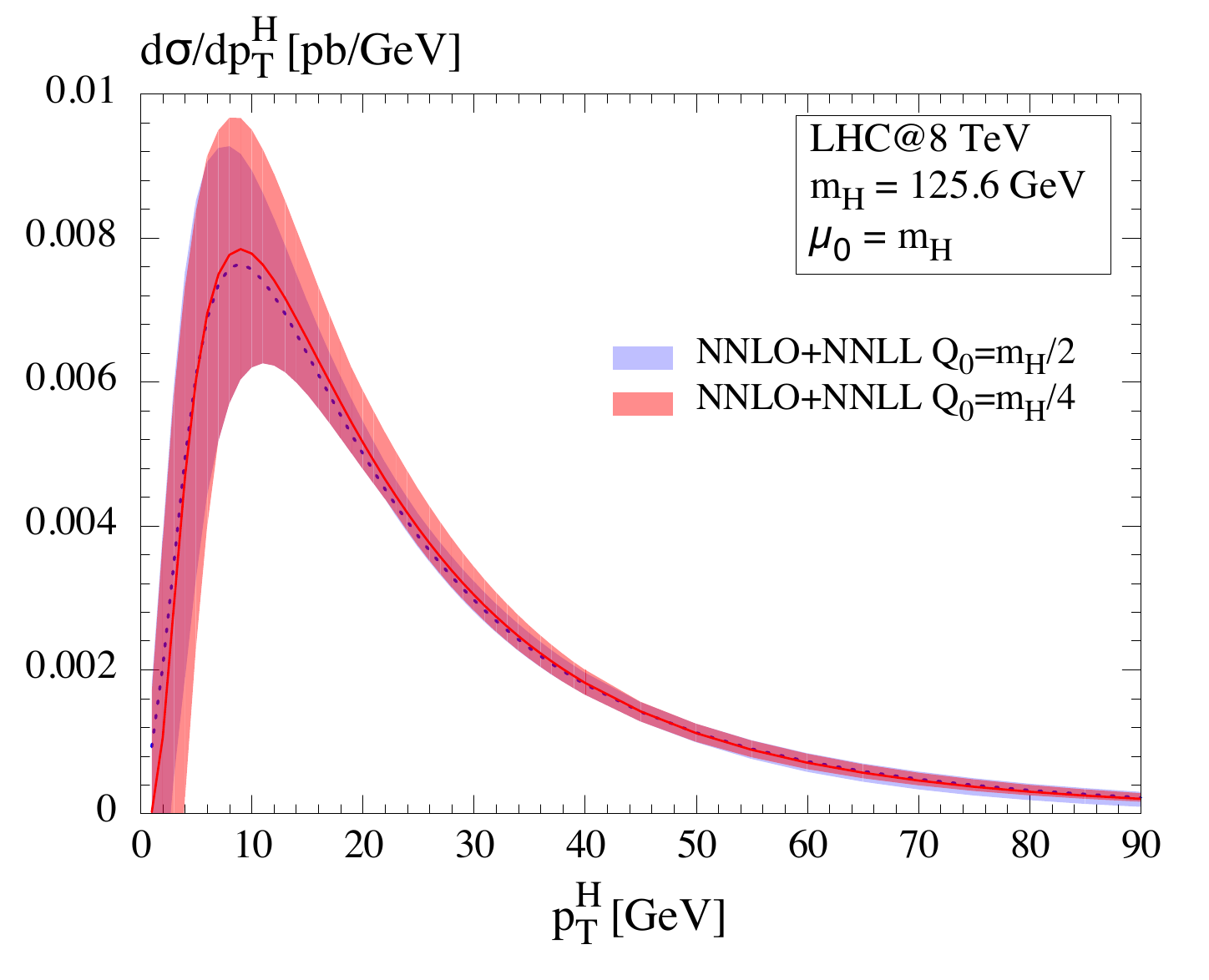}} &
     \mbox{\includegraphics[height=.3\textheight]{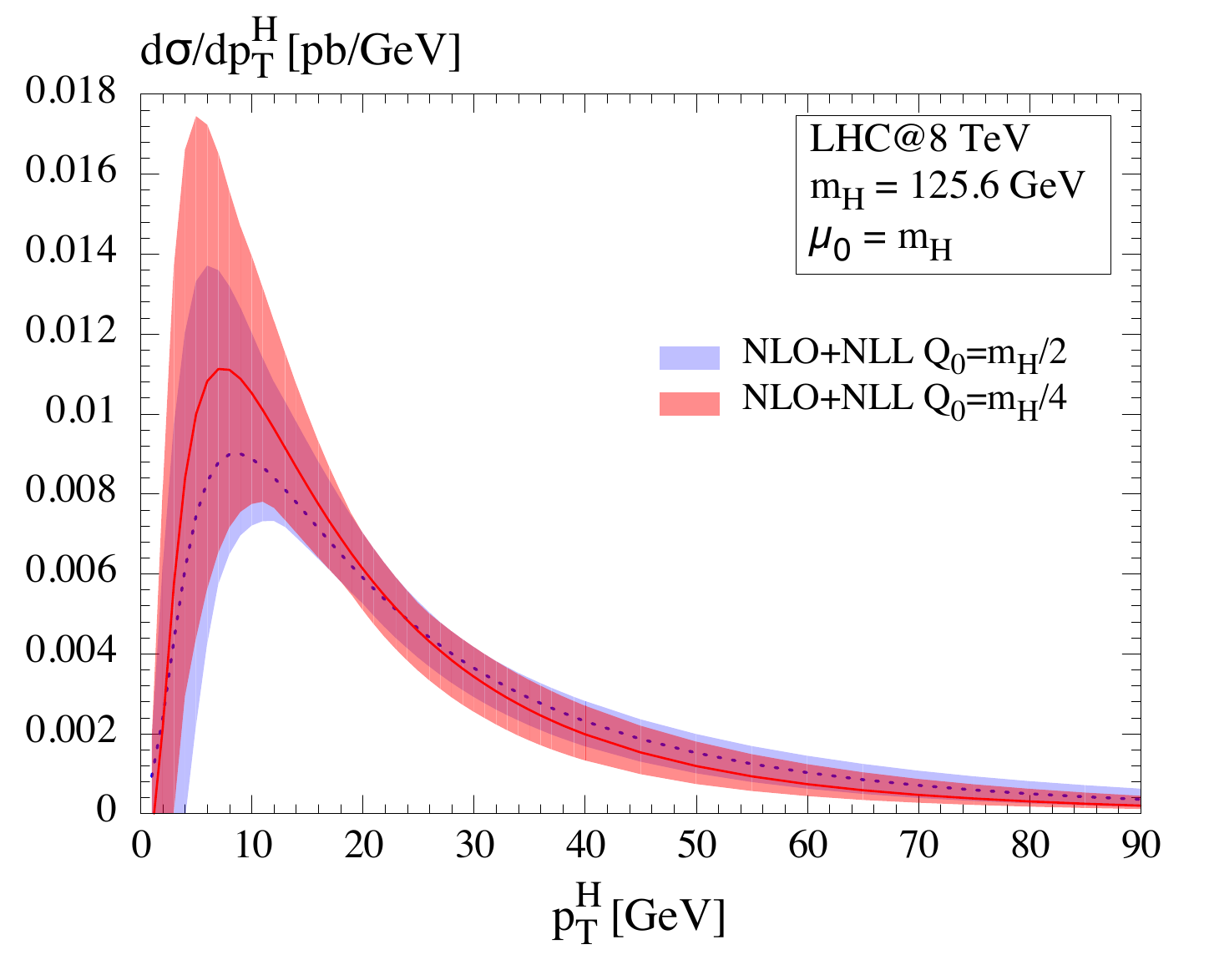}}
            \\
     \hspace{0.15cm} (a) & (b)
    \end{tabular}
    \parbox{.9\textwidth}{%
      \caption[]{\label{fig:QvsQ}{Transverse momentum spectrum using
          $Q_0=M/4$ (red, solid line) and $Q_0=M/2$ (blue, dashed line) as the
          central resummation scale. The bands indicate the theoretical
          uncertainty of the prediction as in \fig{fig:all}. (a)
          \nnlo{}\plus{}\nnll{}; (b) \nlo\plus\nll. } }}
\end{center}
\end{figure}

Let us now consider the effect of the higher orders on the dependence due to
the renormalization and the factorization scale, while fixing the
resummation scale at its default value, $\Qres=Q_0$.  The bands in
\fig{fig:muFmuR} correspond to an independent variation of $\muF$ and
$\muR$ in the range $[\mu_0/2,2\,\mu_0]$, while excluding the region
where $\muF/\muR > 2$ and $\muF/\muR < 1/2$. Comparing the red
\nnlo{}\plus{}\nnll{} with the blue \nlo{}\plus{}\nll{} band, a
considerable decrease of the scale uncertainties is only observed for
$p_T\gtrsim 20$ GeV, while in the region where resummation is crucial
the error bands have a similar size.

Including higher orders in the logarithmic accuracy, one also expects a
reduction of the dependence of the $\pt$ distribution on the resummation
scale. This is impressively confirmed in \fig{fig:Qres}, which shows the
cross sections at \nlo{}\plus{}\nll{} and at
\nnlo{}\plus{}\nnll{}, where $\muF$ and $\muR$ are fixed at
their default values (see \sct{sec:input}). The bands are obtained by
varying $\Qres$ between $Q_0/2$ and $2\,Q_0$; the lines correspond to
$\Qres=Q_0$.  The variation of the cross section with respect to $\Qres$
at \nnll{} is indeed significantly reduced with respect to \nll{}.

Finally, \fig{fig:all} shows the result for an independent variation of
all three scales within $\Qres\in[Q_0/2,2\,Q_0]$ and
$\muF,\muR\in[\mu_0/2,2\,\mu_0]$, where again we exclude the regions
$\muF/\muR > 2$ and $\muF/\muR < 1/2$. For all values of $\pt$, one
observes a reduction of the uncertainty of the resummed-matched
\nnlo{}\plus{}\nnll{} cross section with respect to the one at
\nlo{}\plus\nll{}. The relative uncertainty at the maximum
amounts to $+23/-23$\% for the \nnll{} curve and $+48/-41\%$
at \nll{}.

The corresponding plots for $13$ TeV  are shown in \ref{app:results13},
\fig{fig:all13}-\ref{fig:muFmuR13}. Qualitatively, the above statements 
also apply here, only the absolute cross section is larger.

At this point, we would like to get back to the central choice of the
resummation scale. As we have argued before, $Q=M/4$ results in a good
agreement in the high-$p_T$ tail of the resummed-matched and fixed-order
distribution, particularly at \nnlo{}\plus{}\nnll{}.  Furthermore, at
this order, the choice of $Q_0$ only has a small impact on the
distribution and the corresponding scale uncertainties at low
$p_T$. This is shown in \fig{fig:QvsQ}\,(a), which compares the curves
for $Q_0=M/2$ (blue, dotted line) and $Q_0=M/4$ (red, solid line) at
\nnlo{}\plus{}\nnll{}.  The bands correspond to the variation of all
scales as described above. The height of the peak differs
only by about 3\% between the two choices.
The corresponding curves at
\nlo{}\plus{}\nll{} show a significantly larger difference, which is about 23\% at the peak, see
\fig{fig:QvsQ}\,(b). 
Let us note again that, for further comparison,
plots for $Q_0=M/2$ are given in \ref{app:Q05},
\fig{fig:highpTQ05}-\ref{fig:muFmuR13Q05}.

\begin{figure}
\begin{center}
    \begin{tabular}{cc}
     \hspace{-0.45cm} \mbox{\includegraphics[height=.3\textheight]{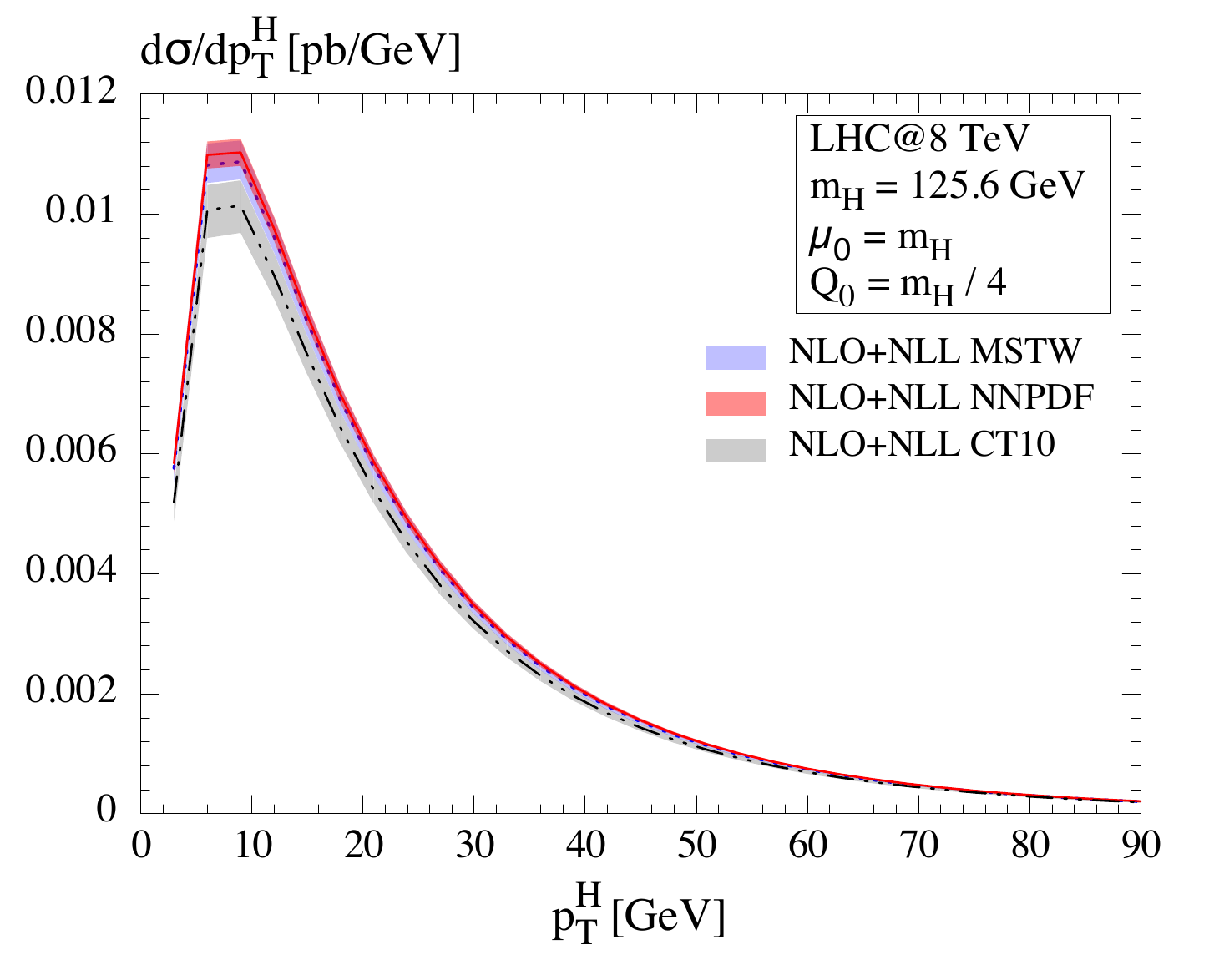}} & \mbox{\includegraphics[height=.3\textheight]{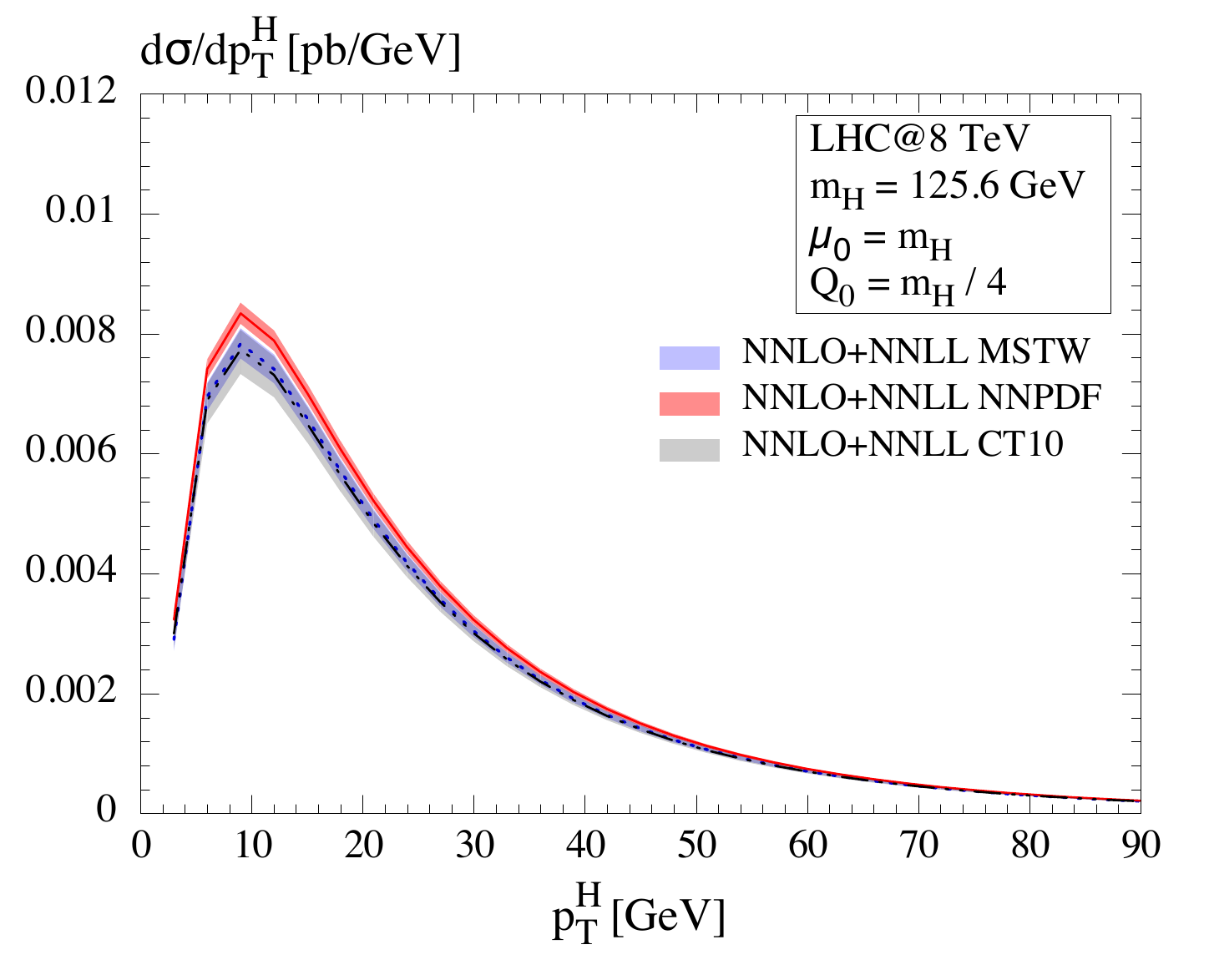}}
            \\
     \hspace{0.15cm} (a) & (b)
    \end{tabular}
    \parbox{.9\textwidth}{%
      \caption[]{\label{fig:pdf_abs}{Resummed-matched
          $p_T$ distribution at (a) \nlo{}\plus\nll{} and (b)
          \nnlo{}\plus\nnll{} for \mstw{} (blue, dotted line), \nnpdf{} (red,
          solid line) and \cteq{} (black, dash-dotted line); lines: central
          curves; bands: $\pdf{}$+$\alpha_s$ uncertainties at 68\% CL.} }}
\end{center}
\end{figure}

\begin{figure}
\begin{center}
\includegraphics[height=.4\textheight]{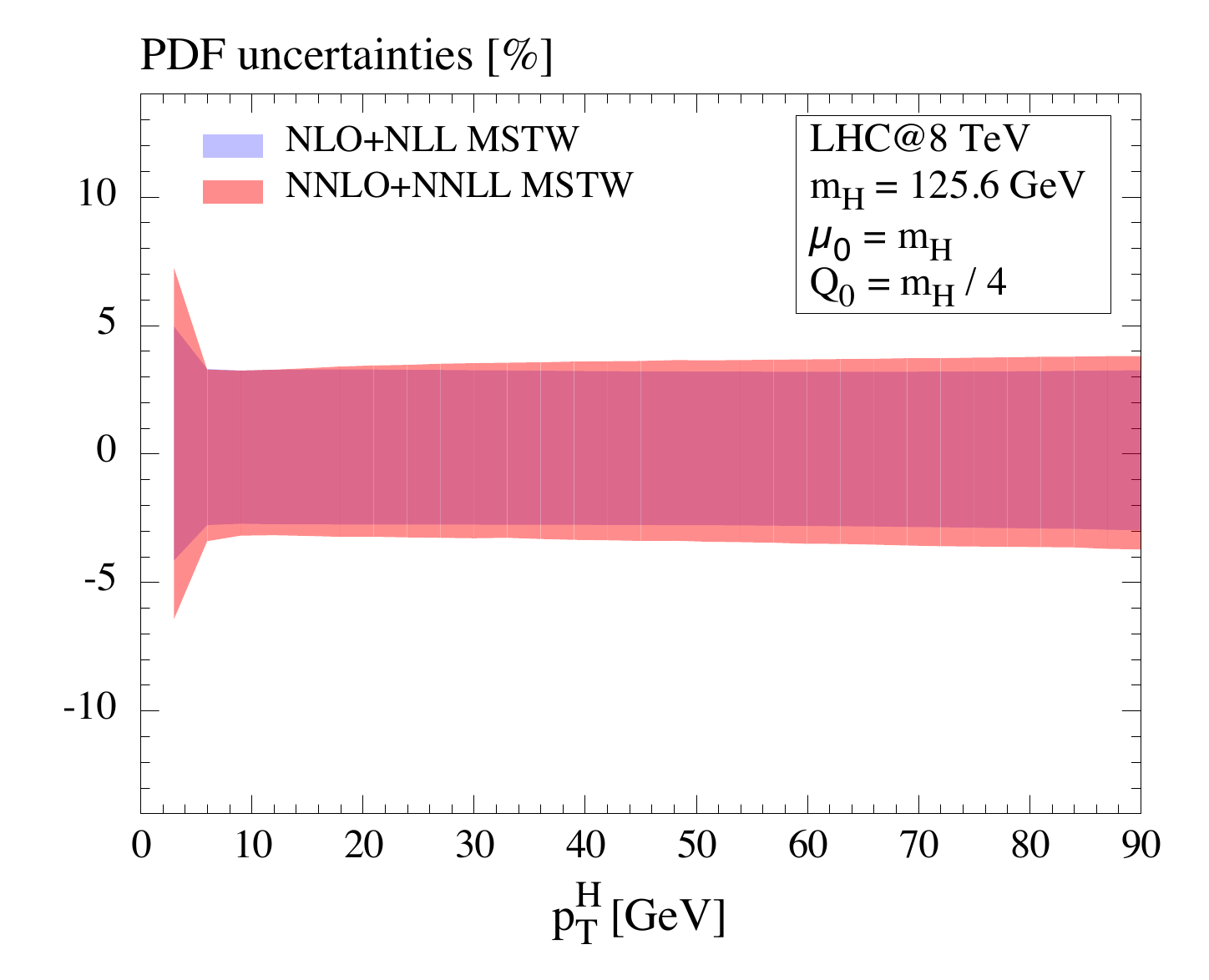}
    \parbox{.9\textwidth}{%
      \caption[]{\label{fig:MSTW}{Relative uncertainties (68\% CL) for the
          \mstw{} \pdf{} set of the resummed-matched $p_T$ distribution
          at \nlo{}\plus\nll{} (blue band) and \nnlo{}\plus\nnll{}
          (red band).  } }}
\end{center}

\begin{center}
\includegraphics[height=.4\textheight]{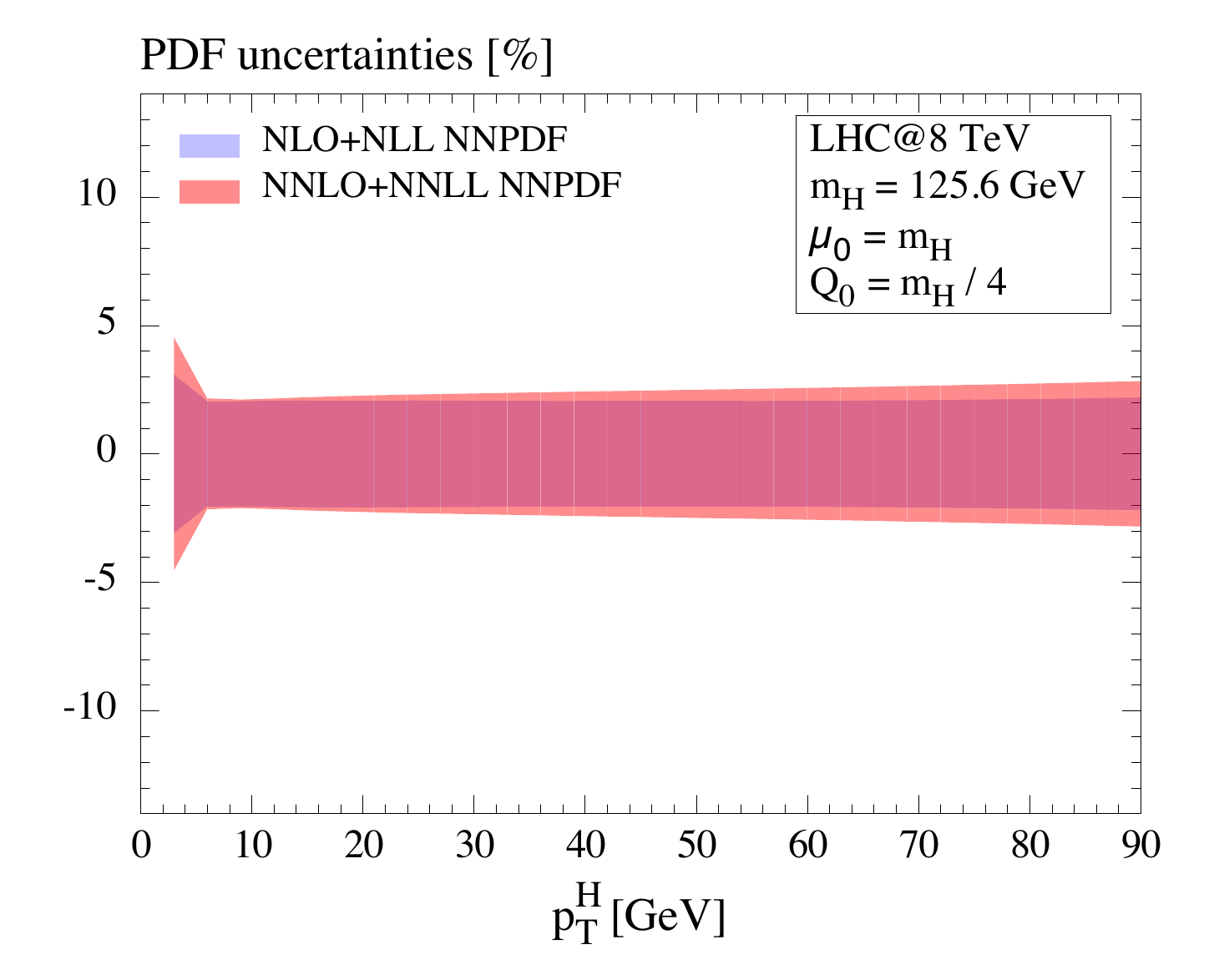}
    \parbox{.9\textwidth}{%
      \caption[]{\label{fig:NNPDF}{Relative uncertainties (68\% CL) for the
          \nnpdf{} \pdf{} set of the resummed-matched $p_T$ distribution
          at \nlo{}\plus\nll{} (blue band) and \nnlo{}\plus\nnll{}
          (red band).  } }}
\end{center}
\end{figure}

\begin{figure}[h]
\begin{center}
\includegraphics[height=.4\textheight]{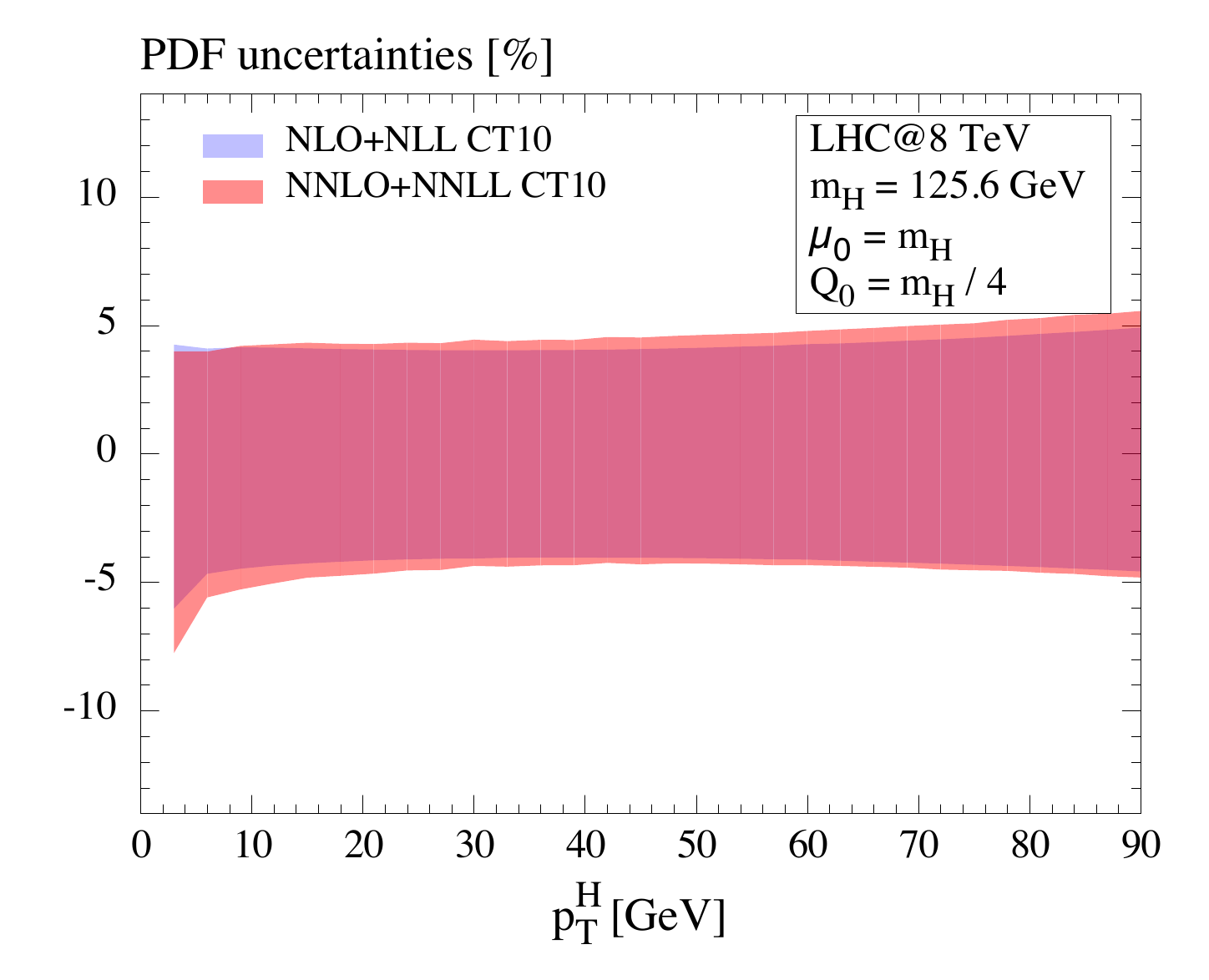}
    \parbox{.9\textwidth}{%
      \caption[]{\label{fig:CTEQ}{Relative uncertainties (68\% CL) for the \cteq{} \pdf{} set of the
         resummed-matched $p_T$ distribution at \nlo{}\plus\nll{} (blue band) and 
         \nnlo{}\plus\nnll{} (red band).
        }
      }}
\end{center}
\end{figure}

Finally, let us discuss the uncertainties arising
from the \pdf{} and $\alpha_s$ choices. Besides the importance for our
calculation, this study is particularly interesting regarding the
treatment of the bottom densities of the various \pdf{} groups, given
the fact that the $\bbh{}$ process in the \fs{5} is directly sensitive
to the bottom densities. We consider three different \pdf{} sets:
\mstw{}, \nnpdf{} and \cteq{}. The combined \pdf{}+$\alpha_s$
uncertainties are determined following the recommendations of the
corresponding \pdf{}
groups~\cite{Martin:2009iq,Ball:2012cx,Lai:2010vv}. In contrast
to {\abbrev MSTW} and {\abbrev CTEQ}, there is no central \pdf{} set for
{\abbrev NNPDF}, which is why the central value is calculated as the
mean value of all considered \pdf{} members.

\fig{fig:pdf_abs} compares the resummed-matched distributions obtained
with the three \pdf{} sets and their intrinsic uncertainties, for
(a)~\nlo{}\plus{}\nll{} and (b)~\nnlo{}\plus{}\nnll{} accuracy. At
\nlo{}\plus{}\nll{}, the {\abbrev MSTW} and {\abbrev NNPDF} results are
very consistent within their uncertainties, while the {\abbrev
  CTEQ} band is right below the {\abbrev MSTW} band. At
\nnlo\plus\nnll{}, on the other hand, the situation is the other way
round: The bands of {\abbrev MSTW} and {\abbrev CTEQ} overlap, while the
{NNPDF} band lies right on top of them. In both cases the biggest
discrepancies are observed around the maximum of the distribution. This
property may be due to the rather special role of the bottom densities
which are not determined directly from experimental data, but are
theoretically derived from the other parton densities and thus, are
strongly dependent on their specific treatment in the different \pdf{}
groups.  Furthermore, considering the relative uncertainties in
\fig{fig:MSTW} ({\abbrev MSTW}), \fig{fig:NNPDF} ({\abbrev NNPDF}) and
\fig{fig:CTEQ} ({\abbrev CTEQ}), we observe \pdf{}+$\alpha_s$
uncertainties of similar size for the \nlo{} and \nnlo{} densities. This
is expected since the \pdf{} uncertainties arise only from the
  experimental input data. In fact, the \nnlo{}
uncertainties are slightly increased with respect to the \nlo{}
ones.  In general the uncertainties of both cross sections
\nlo{}\plus\nll{} and \nnlo{}\plus\nnll{} are rather small, $\lesssim
4$\%, $\lesssim 3$\% and $\lesssim 5$\% for {\abbrev MSTW}, {\abbrev
  NNPDF} and {\abbrev CTEQ}, respectively.

The overall theoretical uncertainty on the cross section is clearly
dominated by unphysical scales, in particular the factorization and
renormalisation scale. It is therefore convenient to simply add the
\pdf{}+$\alpha_s$ and scale uncertainty in quadrature.

\section{Conclusions}
\label{sec:conclusions}
The transverse momentum distribution of Higgs bosons produced in bottom
quark annihilation has been presented through \nnlo\plus\nnll{}
accuracy, following the method of \citere{Bozzi:2005wk}. 
For this purpose, we calculated the missing second-order hard coefficient
$\hardh{,(2)}$ both numerically and analytically. 
By choosing an
appropriate resummation scale, we obtain a resummed-matched distribution
that matches well to the fixed-order prediction at large $\pt$ already
at \nlo\plus\nll. At \nnlo\plus\nnll, we observe excellent agreement
between the resummed-matched and the fixed-order curve already above
around 50~GeV. Our results therefore represent a precise prediction in
the dominant region of low and intermediate values of transverse
momenta.

Concerning the variation of the cross section with the unphysical
scales, we observe a significant reduction when going from
\nlo{}\plus\nll{} to \nnlo{}\plus\nnll{}. In fact, the extremely weak
dependence of the \nnlo{}\plus\nnll{} result on the resummation scale is
remarkable. 
The \pdf{} uncertainties are roughly of the same size
at \nlo\plus\nll{} and \nnlo\plus\nnll, as it is expected from their
purely experimental origin.

Our results should prove useful, in particular, in scenarios with enhanced
bottom quark Yukawa coupling, such as supersymmetric or Two-Higgs-Doublet
Models with large values of $\tan\beta$, but they may even be an
important complement in the \sm{}, especially once the statistics for
Higgs events has improved. 
Not least of all, differential quantities in the
$\bbh$ process provide good physical observables to study various
parametrizations and implementations of $b$ densities.

\paragraph{Acknowledgements.}
We would like to thank Giancarlo Ferrera and Massimiliano Grazzini for
enlightening discussions, and J.\,Bl\"umlein for helpful
communication. The work of A.T. was supported in part by the University of
Torino, the Compagnia di San Paolo, contract ORTO11TPXK, and DFG,
contract HA\,2990/5-1. M.W. was supported by BMBF, contract 05H12PXE, and
by the European Commission through the FP7 Marie Curie Initial Training
Network ``LHCPhenoNet'' (PITN-GA-2010-264564).

\appendix
\gdef\thesection{Appendix \Alph{section}}
\newpage
\section{Feynman diagrams}
\label{app:diag}
\begin{figure}[h]
  \begin{center}
    \begin{tabular}{ccc}
     \hspace{0.4cm} \mbox{\includegraphics[height=.125\textheight]{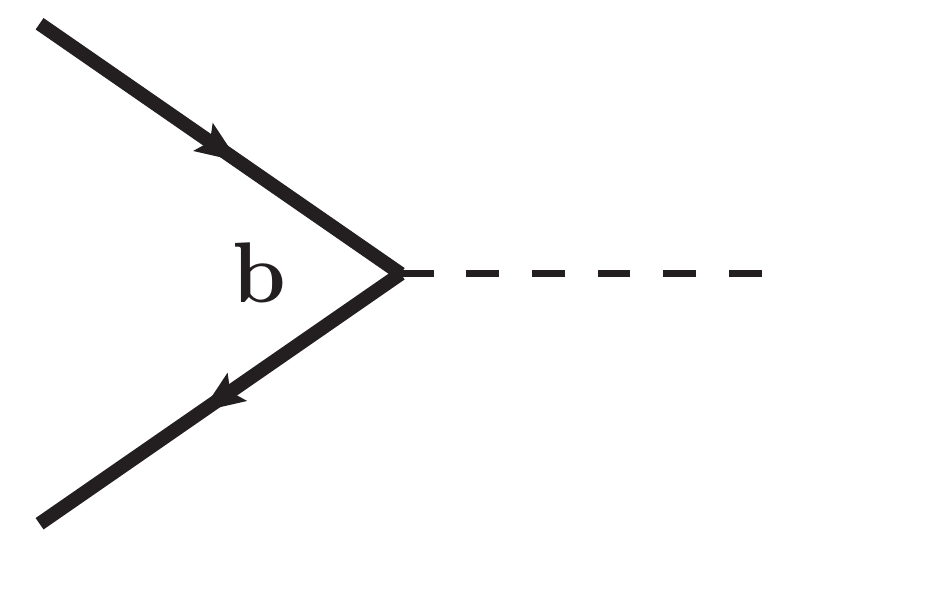}} & \mbox{\includegraphics[height=.125\textheight]{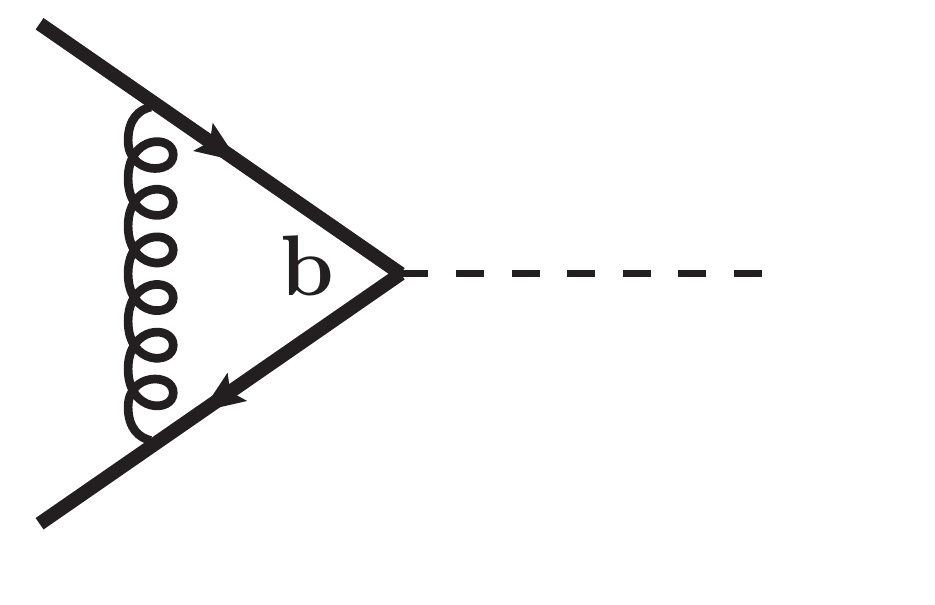}} &
      \mbox{\includegraphics[height=.125\textheight]{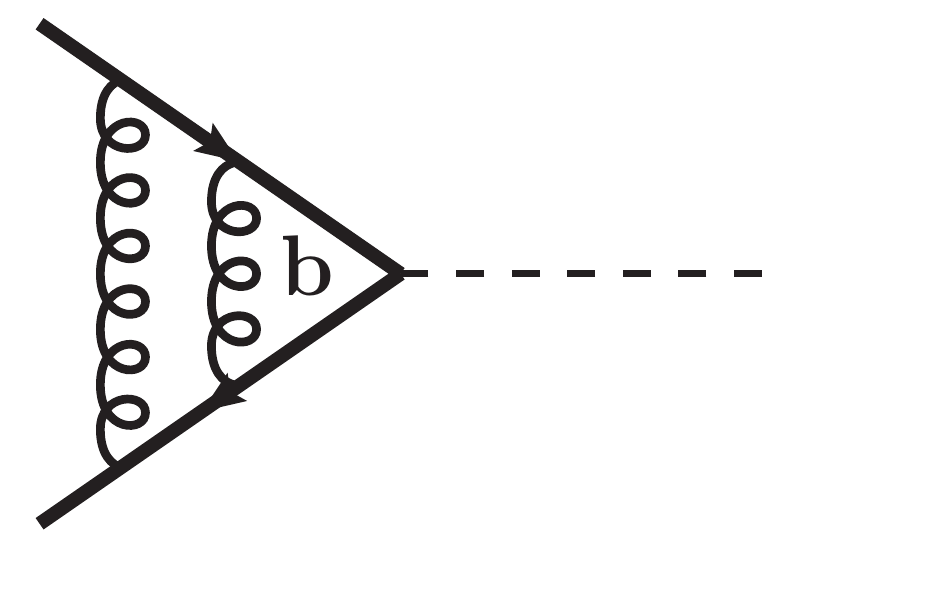}}
            \\
     \hspace{0.15cm} (a) & (b) & (c)
    \end{tabular}
    \parbox{.9\textwidth}{%
      \caption[]{\label{fig:app1f2}
        A sample of Feynman diagrams for $b\bar{b}\rightarrow H$ contributing to the \nnlo{} cross section at $\pt=0$\,; (a) \lo{}, (b) one-loop and (c) two-loop.
        }
      }
  \end{center}
\end{figure}
\begin{figure}[h]
  \begin{center}
    \begin{tabular}{ccccc}
      \mbox{\includegraphics[height=.11\textheight]{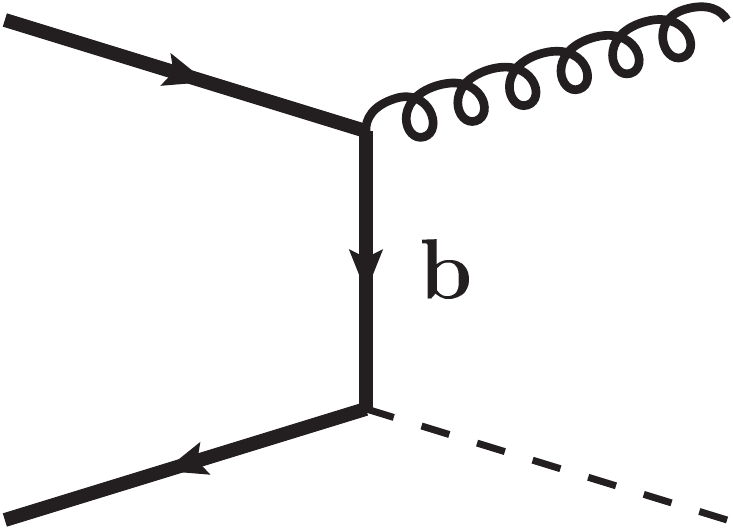}} & & \mbox{\includegraphics[height=.11\textheight]{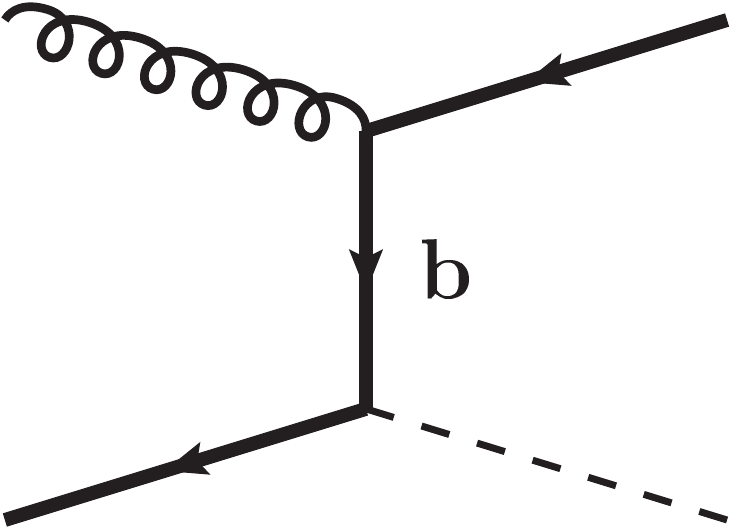}} & &
      \mbox{\includegraphics[height=.11\textheight]{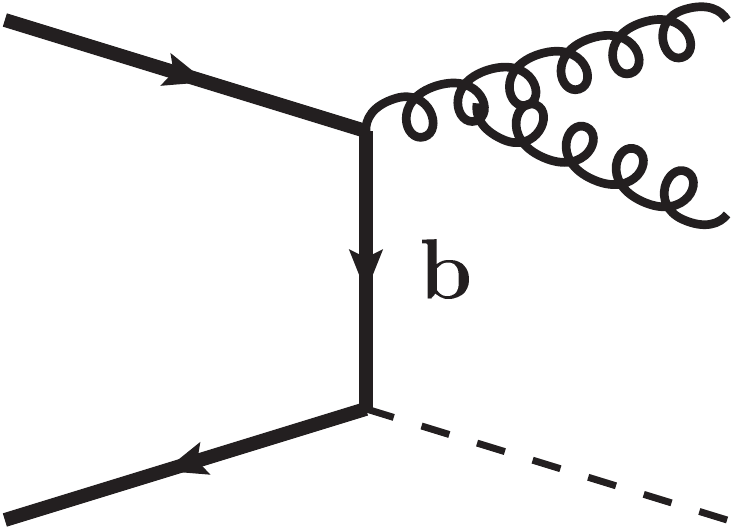}}
            \\
      (a) & & (b) & & (c)\\
      \mbox{\includegraphics[height=.12\textheight]{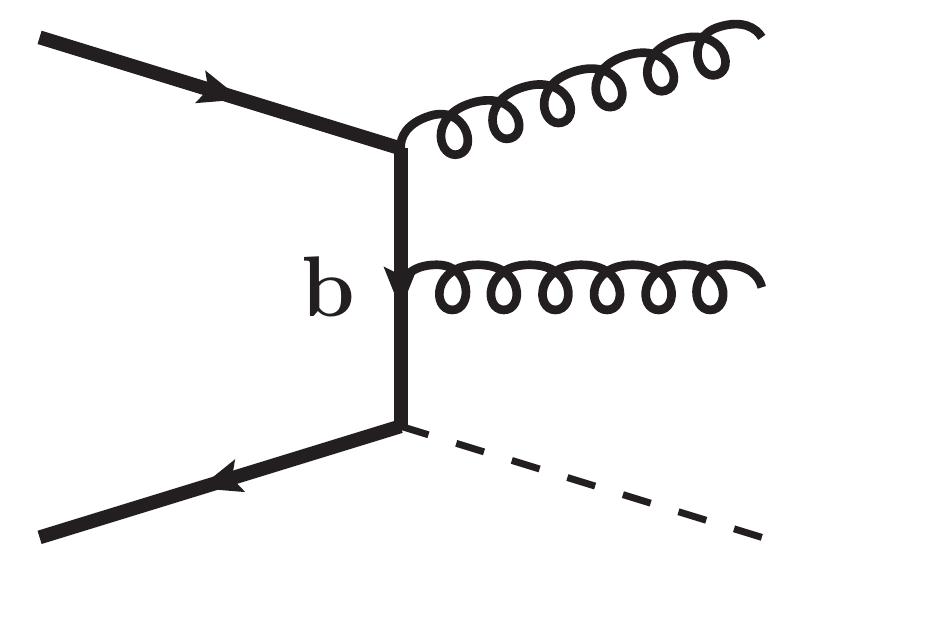}} & & \mbox{\includegraphics[height=.12\textheight]{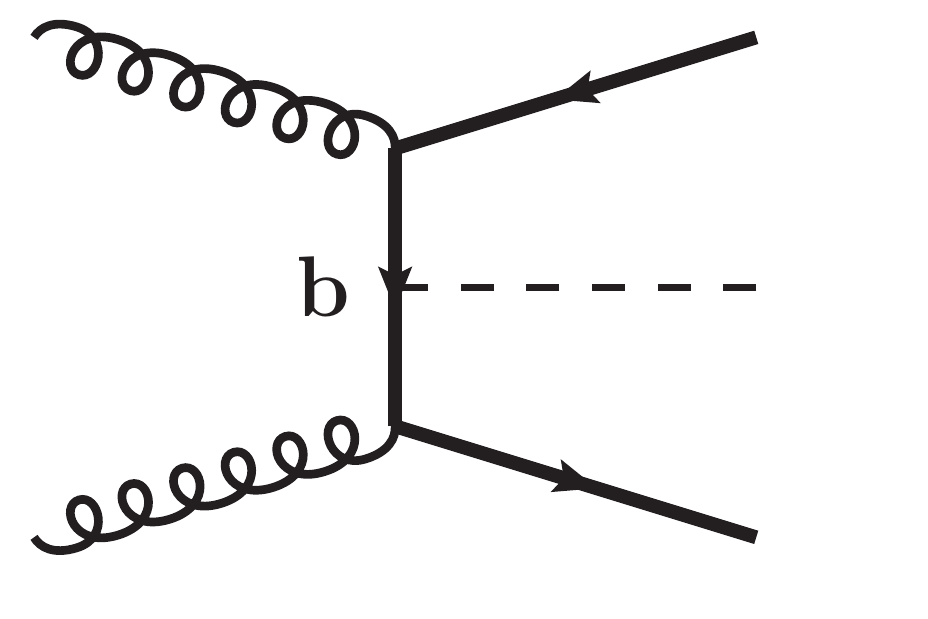}} & &
      \mbox{\includegraphics[height=.11\textheight]{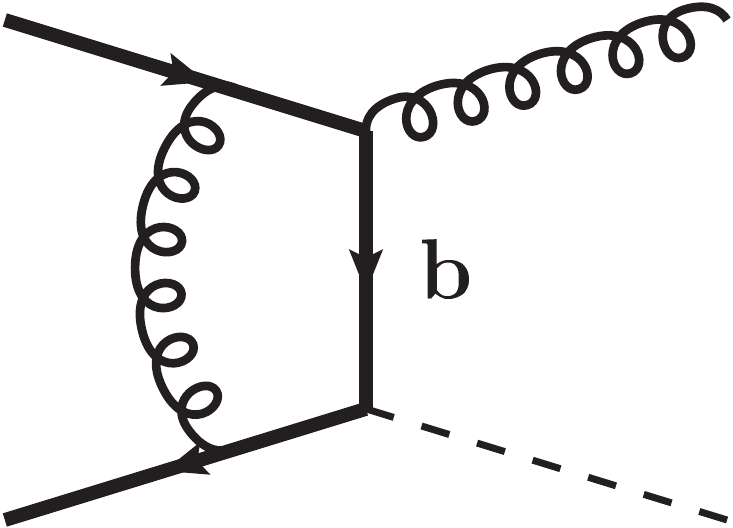}}
            \\
      (d) & & (e) & & (f)
    \end{tabular}
    \parbox{.9\textwidth}{%
      \caption[]{\label{fig:app1f1}
        A sample of Feynman diagrams for $b\bar{b}\rightarrow H$ contributing to the \nnlo{} cross section at $\pt>0$\,; (a-b) single-real, (c-e) double-real, (f) mixed real-virtual.
        }
      }
  \end{center}
\end{figure}

\section{Hard-collinear coefficient with full scale dependence}

In this appendix, we present expressions for the hard-collinear function
to second order with complete scale dependence for the $\bbh$
process:
\begin{align}
\begin{split}
\chardh{,(1)}{,ij}(z)&=\delta(1-z)\,\delta_{bi}\,\delta_{\bar{b}j}\,
\left[\hardh{,(1)}
-\left(B_b^{(1)}+\frac12\,A_b^{(1)}\right)-2\,\gamma_0\,\lnR\right]
\\ &+\delta_{bi}\,C_{\bar{b}j}^{(1)}(z)+\delta_{\bar{b}j}\,C_{bi}^{(1)}(z)+\frac12\left(\delta_{bi}\,P_{\bar{b}j}^{(0)}(z)+\delta_{\bar{b}j}\,P_{bi}^{(0)}(z)\right)\,\lnthreeQ\,,
\end{split}
\end{align}

\begin{align}
\begin{split}
\label{eq:cH2}
\chardh{,(2)}{,ij}&(z)=\delta(1-z)\,\delta_{bi}\,\delta_{\bar{b}j}\,
\hardh{,(2)}+\delta_{bi}\,C_{\bar{b} j}^{(2)}(z)+\delta_{\bar{b}
  j}\,C_{bi}^{(2)}(z)+(C_{bi}^{(1)}\otimes C_{\bar{b} j}^{(1)})(z)
\\
&+H_b^{h(1)}\left(\delta_{bi}\,C_{\bar{b}j}^{(1)}(z)+\delta_{\bar{b}j}\,C_{bi}^{(1)}(z)\right)+\delta(1-z)\,\delta_{bi}\,\delta_{\bar{b}j}\,\frac16\,A_b^{(1)}\,\beta_0\,\lnthreeQ
\\
&+\frac12\left[\delta(1-z)\,\delta_{bi}\,\delta_{\bar{b}j}\,A^{(2)}_b+\beta_0\,\Sigma_{\bbbar\leftarrow
    ij}^{(1;1)}(z)\right]\,\lntwoQ\\ 
&-\bigg[\delta(1-z)\,\delta_{bi}\,\delta_{\bar{b}j}\,\left(B^{(2)}_b+A^{(2)}_b
  \lnQ\right)
  \\ &-\beta_0\left(\delta_{bi}\,C_{\bar{b}j}^{(1)}(z)+\delta_{\bar{b}j}\,C_{bi}^{(1)}(z)\right)+\delta_{bi}\,\frac14
  P_{\bar{b}j}^{(1)}(z)+\delta_{\bar{b}j}\,\frac14
  P_{bi}^{(1)}(z)\bigg]\,\lnQ \\
&+\frac14\beta_0\left(\delta_{bi}\,P_{\bar{b}j}^{(0)}(z)+\delta_{\bar{b}j}\,P_{bi}^{(0)}(z)\right)\lntwoF
\\
&+\frac14\left(\delta_{bi}\,P_{\bar{b}j}^{(1)}(z)+\delta_{\bar{b}j}\,P_{bi}^{(1)}(z)\right)
\lnF-\chardh{,(1)}{,ij}(z)\,\beta_0\lnR \\
&+\frac12\sum\limits_{i',j'}\Bigg[\chardh{,(1)}{,i'j'}(z)+\delta(1-z)\,\delta_{bi'}\,\delta_{\bar{b}
    j'}\,\hardh{,(1)}+\delta_{bi'}\,C^{(1)}_{\bar{b}
    j'}(z)+\delta_{\bar{b} j'}\,C^{(1)}_{bi'}(z)\Bigg] \\
&\times\Bigg\{\frac12\left(\delta_{i'i}\,P_{j'j}^{(0)}(z)+\delta_{j'j}\,P_{i'i}^{(0)}(z)\right)\lnQF-\delta(1-z)\,\delta_{i'i}\,\delta_{j'j}
\\
&\times\bigg[\left(B^{(1)}_b+\frac12\,A^{(1)}\lnQ\right)\lnQ+2\,\gamma_0\lnR\bigg]\Bigg\}
\\ &-\delta(1-z)\,\delta_{bi}\,\delta_{\bar{b}
  j}\left[\gamma_0\,\beta_0\lntwoR+2\gamma_1\lnR\right],
\end{split}
\end{align}
where $M$ denotes the Higgs mass, $\Sigma_{\bbbar\leftarrow ij}^{(1;1)}$
is defined in Eq.\,(64) of \citere{Bozzi:2005wk}, and $P_{ij}^{(n)}(z)$
denotes the Altrelli-Parisi splitting functions.
Their expressions can be found in \citere{Ellis:1996nn}, for example.
The quark mass anomalous dimension enters due to the fact that the Born
factor is proportional to the square of the bottom quark mass (see
\eqn{eq:born}) which is normalized in the $\msbar{}$ scheme:
\begin{equation}
\begin{split}
\gamma_0 &= \frac34\,C_F, \\
\gamma_1 &= \frac{1}{16} \left(   
\frac{3}{2}\,C_F^2 + \frac{97}{6}C_F\,C_A - \frac{10}{3} \,C_F\, T_F \,N_f
\right),
\end{split}
\end{equation}
while the power of $\alpha_s$ at \lo{} vanishes.

\allowdisplaybreaks
\newpage
\section{Mellin transforms}
\label{app:mellins}
Mellin transforms of several transcendental functions which appear in
two-loop calculations are reported for integer $N$ in
\citere{Blumlein:1998if}.  \citere{Blumlein:2000hw} gives a {\abbrev
  FORTRAN} code that numerically approximates the analytic continuation
of the moments of 25 basic functions\footnote{The formula for $g(11,N)$ in Eq.\,(30) of
  \citere{Blumlein:2000hw} contains a typo: The last term $\frac{1}{4}
  \ln^4 2$ should be replaced by $\frac{1}{8} \ln^4 2$. We would like to
  thank J.\,Bl\"umlein for confirmation.} termed
$g(1,z), \dots, g(25,z)$ (see Section 3 of \citere{Blumlein:2000hw}). The
resummation coefficients $C^{(1)}_N$ and $C^{(2)}_N$ can be expressed in
terms of the moments of these 25 basic functions $g(1,N), \dots, g(25,N)$, and the analytic
continuation of the single harmonic sums $S_k(N)$.
Below, we give analytic expressions for Mellin transforms 
defined by
\begin{equation}
f(N) = \int_0^1 ~dz z^{N-1} f(z)
\end{equation}
of some of the transcedental functions, true for 
complex $N$, which appear in the coefficients $C^{(2)}$ of our calculation. 
The general definition of the
harmonic sums is given by
\begin{align}
S_{k_1, \ldots. k_m}(N) = 
\sum_{n_1 =1}^{N}
\frac{\left(\text{sign}(k_1) \right)^{n_1}} {n_1^{|k_1|}}
\cdots
\sum_{n_m =1}^{n_m-1}
\frac{\left(\text{sign}(k_m) \right)^{n_m}} {n_m^{|k_m|}},
\end{align}
which are defined, of course, only for integer $N$ and for $k_i \neq 0$. The analytic continuations are 
known for single sums and are expressed in terms of 
the digamma function
$ \psi(N) =\psi^{(0)}(N)$
and the polygamma functions
$\psi^{(m)}(N)$. They are given by
\begin{align}
\begin{split}
S_{k}(N) &= (-1)^{k-1}\frac{1}{(k-1)!} \psi^{(k-1)}(N+1) + c_k^{+},
\\ 
S_{-k}(N) &= (-1)^{k-1+N}\frac{1}{(k-1)!} \beta^{(k-1)}(N+1) - c_k^{-},
\label{beta}
\end{split}
\end{align}
where
\begin{align}
\begin{split}
\psi(N) &= \frac{1}{\Gamma(N)} \frac{d \Gamma(N)}{dN}, \hspace{3.17cm} \psi^{(m)}(N) = \frac{d^{m} \psi(N)}{dN^m},\\
\beta(N) &= \frac{1}{2} \left[   
\psi \left( \frac{N+1}{2} \right)
- \psi \left( \frac{N}{2} \right)
\right],\quad\quad \beta^{(m)}(N) = \frac{d^{m} \beta(N)}{dN^m}
\end{split}
\end{align}
and 
\begin{align}
\begin{split}
c_1^{+} &= \gamma_E,  \hspace{1cm} c_k^+ = \zeta_k \equiv \zeta(k), \quad k \geq 2,
\\
c_1^{-} &= \log2,  \quad \quad c_k^{-} = \left(1- \frac{1}{2^{k-1}} \right) \zeta_k, \quad k \geq 2.
\end{split}
\end{align}

We obtained the following Mellin transforms by modifying some of the formulas in \citere{Blumlein:1998if,Blumlein:2000hw} such that they become valid for complex $N$:
\begin{align}
\begin{split}
\ln(1+z) \; &\rightarrow \;\;
\frac{1}{2N} \left[ S_1\left( \frac{N-1}{2} \right)
-  S_1\left( \frac{N}{2} \right)
+ 2 \log 2 \right]
\end{split}
\end{align}
\begin{align}
\begin{split}
\Li_2(-z)  \;& \rightarrow \;\; 
 \frac{1}{2N^2} 
\left[ 
S_1 \left( \frac{N-1}{2} \right)
-S_1 \left( \frac{N}{2} \right)
+2 \log2
\right]
-\frac{\zeta_2}{2N}
\end{split}
\end{align}
\begin{align}
\begin{split}
\ln(z) \ln(1+z) \;& \rightarrow \;\; 
-\frac{1}{2N^2} 
\left[ 
S_1 \left( \frac{N-1}{2} \right)
-S_1 \left( \frac{N}{2} \right)
+2 \log2
\right]
\\
&
\hspace{0.77cm}
-\frac{1}{4N} 
\Bigg[ 
S_2 \left( \frac{N-1}{2} \right)-S_2 \left( \frac{N}{2} \right)
\Bigg]
\end{split}
\end{align}

\begin{align}
\begin{split}
\Li_3 \bigg( &\frac{1-z}{1+z}   \bigg)
- \Li_3 \left(- \frac{1-z}{1+z}   \right)
 \; \rightarrow \;\;\left(S_1 \left(\frac{N - 1}{2}\right) - 
     S_1\left(\frac{N}{2}\right) \right)
\\
& 
\times \frac{1}{8N} 
\Bigg[\psi^{(1)} \left(\frac{N + 1}{2}\right)  - 4 \psi^{(1)}(N + 1) - 
     \psi^{(1)} \left(\frac{N + 2}{2} \right) + \pi^2 \Bigg] 
 \\
&
+ \frac{1}{4N} 
\Bigg[\psi^{(1)} \left(\frac{N + 1}{2}\right)  - 4 \psi^{(1)}(N + 1) - 
     \psi^{(1)} \left(\frac{N + 2}{2} \right) \Bigg] 
     \left(S_1\left(N\right) + \ln2 \right)
 \\
&
+\frac{1}{N} \Bigg[g(3,N + 1) - g(4,N + 1) + g(18,N + 1) - g(19,N + 1) \Bigg] ,
\end{split}
\end{align}
\begin{align}
\begin{split}
&\frac{\ln(1+z) \ln^2(z)}{1+z} \; \rightarrow \;\;
2 \, g(5,N)-2 \, g(6,N)-2 \, g(7,N)\\
&\quad\quad +(-1)^{-N} \Bigg[-\frac{1}{2} \zeta_3 S_{-1}(N-1)
 +\frac{3}{2} \zeta_3 S_1(N-1)+4 S_{-4}(N-1)\\
 &\quad\quad -\frac{1}{2} \pi ^2 S_{-2}(N-1)
+2 S_{-3}(N-1) S_1(N-1)+2 S_{-2}(N-1) S_2(N-1)
 \\ &\quad\quad
+\frac{1}{6} \pi ^2 S_2(N-1)-\frac{1}{2}
   \zeta_3 \ln (2)-\frac{\pi ^4}{360}\Bigg],
\end{split}
\label{loglog2}
\end{align}
\begin{align}
\begin{split}
& \frac{1}{1+z} \left[\Li_3\left(\frac{1}{1+z} \right) - \frac{1}{6} \ln^3(1+z) \right]\; \rightarrow  
 \\ 
\hspace{-5cm}
& \frac{1}{192} \Bigg[\psi ^{(0)}\left(\frac{N}{2}\right) \left(-24
   \, g\left(18,\frac{N}{2}\right)+24 \, g(19,N)-9 \zeta_3+\pi ^2 (6 \gamma +\ln
   (16))\right)
 \\ 
\hspace{-5cm}
& 
-\psi ^{(0)}\left(\frac{N+1}{2}\right) \left(-24
   \, g\left(18,\frac{N+1}{2}\right)+24 \, g(19,N)-9 \zeta_3+ \pi ^2 (6 \gamma +\ln
   (16))\right)
 \\ 
\hspace{-5cm}
& 
-4 \pi ^2 \, g(1,N)-144 \, g(5,N)+192 \, g(6,N)+144 \, g(7,N)+96 \, g(8,N)+48
   \, g(10,N)
 \\ 
\hspace{-5cm}
& 
+96 \, g(11,N)+48 \, g(12,N)
+24 \, g\left(20,\frac{N}{2}\right)-24
   \, g\left(20,\frac{N+1}{2}\right)+24 \, g\left(21,\frac{N}{2}\right)
 \\ 
\hspace{-5cm}
& 
-24
   \, g\left(21,\frac{N+1}{2}\right)+24 \bigg(-\gamma 
   \, g\left(18,\frac{N}{2}\right)+\gamma  \, g\left(18,\frac{N+1}{2}\right)+\ln (4)
   \, g(4,N)
 \\ 
\hspace{-5cm}
& 
+4 (\psi ^{(0)}(N)+\gamma ) (g(3,N)-g(4,N))\bigg)+9 \psi
   ^{(1)}\left(\frac{N}{2}\right) \psi ^{(0)}\left(\frac{N}{2}\right)^2
 \\ 
\hspace{-5cm}
& 
   -12 \psi
   ^{(2)}\left(\frac{N}{2}\right) \psi ^{(0)}\left(\frac{N}{2}\right)
+3 \psi
   ^{(1)}\left(\frac{N+1}{2}\right) \psi ^{(0)}\left(\frac{N}{2}\right)^2+6 \pi ^2
   \psi ^{(0)}(N) \psi ^{(0)}\left(\frac{N}{2}\right)
 \\ 
\hspace{-5cm}
& 
+6 \psi
   ^{(0)}\left(\frac{N+1}{2}\right) \psi ^{(1)}\left(\frac{N}{2}\right) \psi
   ^{(0)}\left(\frac{N}{2}\right)+24 \gamma\,  \psi ^{(1)}\left(\frac{N}{2}\right) \psi
   ^{(0)}\left(\frac{N}{2}\right)
 \\ 
\hspace{-5cm}
& 
-6 \psi ^{(0)}\left(\frac{N+1}{2}\right) \psi
   ^{(1)}\left(\frac{N+1}{2}\right) \psi ^{(0)}\left(\frac{N}{2}\right)
   -12 \psi ^{(0)}(N)^2 \psi
   ^{(1)}\left(\frac{N+1}{2}\right)
 \\ 
\hspace{-5cm}
& 
+48 \psi
   ^{(2)}(N) \psi ^{(0)}\left(\frac{N}{2}\right)-6 \pi ^2 \psi ^{(0)}(N) \psi
   ^{(0)}\left(\frac{N+1}{2}\right)+12 \psi ^{(0)}(N)^2 \psi
   ^{(1)}\left(\frac{N}{2}\right)
 \\ 
\hspace{-5cm}
& 
-3 \psi ^{(0)}\left(\frac{N+1}{2}\right)^2 \psi
   ^{(1)}\left(\frac{N}{2}\right)+48 \psi ^{(0)}(N) \psi
   ^{(0)}\left(\frac{N+1}{2}\right) \psi ^{(1)}(N)
 \\ 
\hspace{-5cm}
& 
-9 \psi ^{(0)}\left(\frac{N+1}{2}\right)^2 \psi
   ^{(1)}\left(\frac{N+1}{2}\right)-24 \gamma  \psi ^{(0)}\left(\frac{N+1}{2}\right)
   \psi ^{(1)}\left(\frac{N+1}{2}\right)
 \\ 
\hspace{-5cm}
& 
-48 \psi ^{(0)}\left(\frac{N+1}{2}\right) \psi
   ^{(2)}(N)+12 \psi ^{(0)}\left(\frac{N+1}{2}\right) \psi
   ^{(2)}\left(\frac{N+1}{2}\right)+\psi ^{(3)}\left(\frac{N}{2}\right) 
 \\ 
\hspace{-5cm}
& 
-\psi^{(3)}\left(\frac{N+1}{2}\right)
+6 \left(-\pi ^2-2 \ln ^2(2)+\gamma  (4 \gamma
   +\ln (16))\right) \psi ^{(1)}\left(\frac{N}{2}\right)+6 \Big[\pi ^2
 \\ 
\hspace{-5cm}
& 
+2 \ln
   ^2(2)
-\gamma  (4 \gamma +\ln (16))\Big] \psi ^{(1)}\left(\frac{N+1}{2}\right)
+48
   (\ln (2)-\gamma ) \psi ^{(1)}(N) \psi ^{(0)}\left(\frac{N}{2}\right)
 \\ 
\hspace{-5cm}
& 
+24 (\gamma
   +\ln (2)) \psi ^{(0)}(N) \psi ^{(1)}\left(\frac{N}{2}\right)+48 (\gamma -\ln (2))
   \psi ^{(0)}\left(\frac{N+1}{2}\right) \psi ^{(1)}(N)
 \\ 
\hspace{-5cm}
& 
-24 (\gamma +\ln (2)) \psi
   ^{(0)}(N) \psi ^{(1)}\left(\frac{N+1}{2}\right)-12 (\gamma +\ln (2)) \psi
   ^{(2)}\left(\frac{N}{2}\right)
 \\ 
\hspace{-5cm}
&
+12 (\gamma +\ln (2)) \psi
   ^{(2)}\left(\frac{N+1}{2}\right)-48 \psi ^{(0)}(N) \psi ^{(1)}(N) \psi
   ^{(0)}\left(\frac{N}{2}\right)\Bigg],
\end{split}
\end{align}
where $\gamma =\gamma_E$. Note that in \eqn{loglog2} factors of the form $(-1)^N$ cancel out completely
upon insertion of the single harmonic sums in \eqn{beta} .

\section{Results for \bld{13} TeV}\label{app:results13}

\begin{figure}[h]
\begin{center}
\includegraphics[height=.385\textheight]{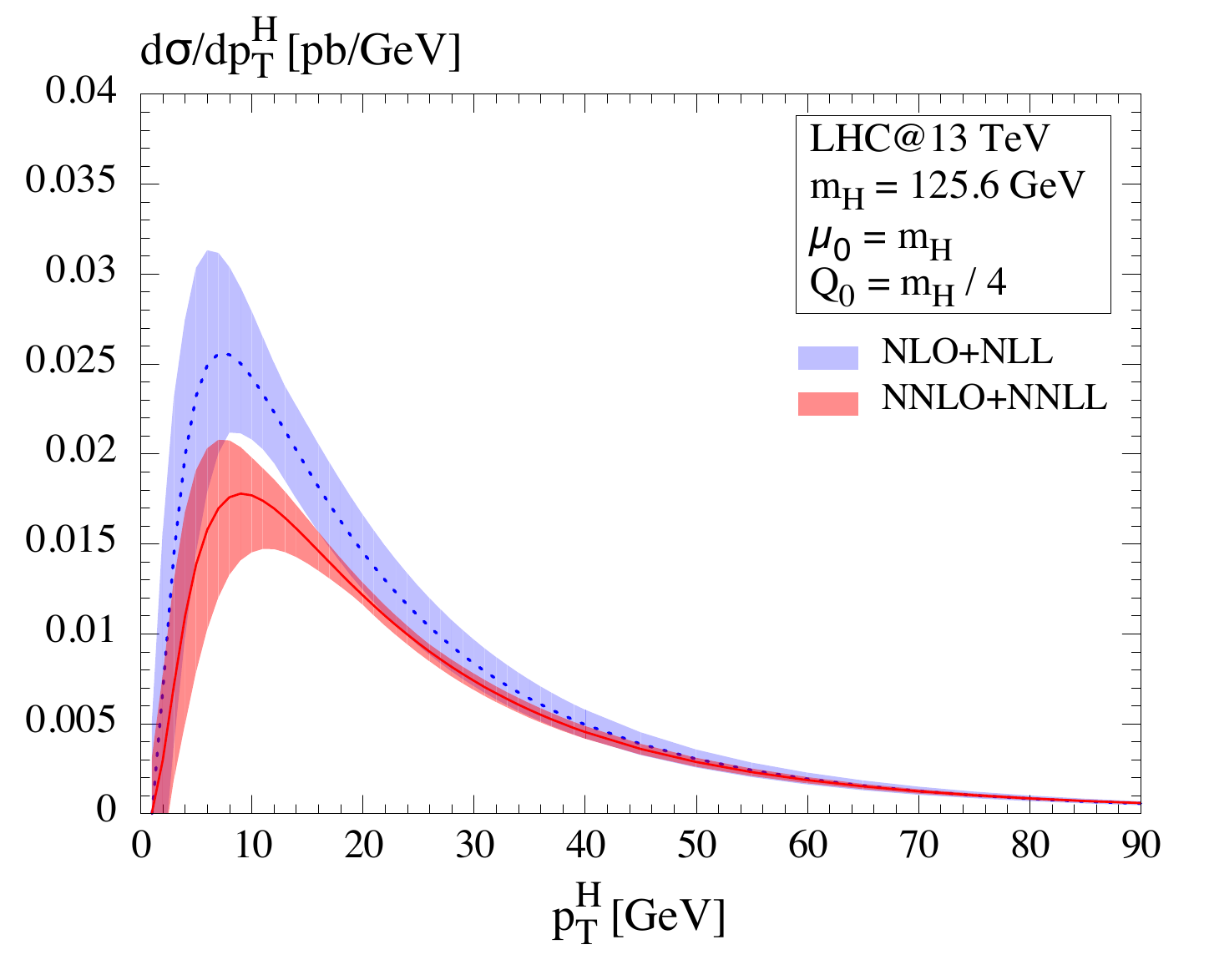}
    \parbox{.9\textwidth}{%
      \caption[]{\label{fig:muFmuR13}{Resummed-matched $p_T$-distribution at \nlo{}\plus\nll{} (blue, dashed line) and \nnlo{}\plus\nnll{} (red, solid line); lines: central scale choices; bands: uncertainty due to $\muF{},\muR{}$-variation. (Same as \fig{fig:muFmuR}, but for $\sqrt{s}=13$\,TeV.)
        }
      }}
\end{center}
\end{figure}

\begin{figure}
\begin{center}
\includegraphics[height=.385\textheight]{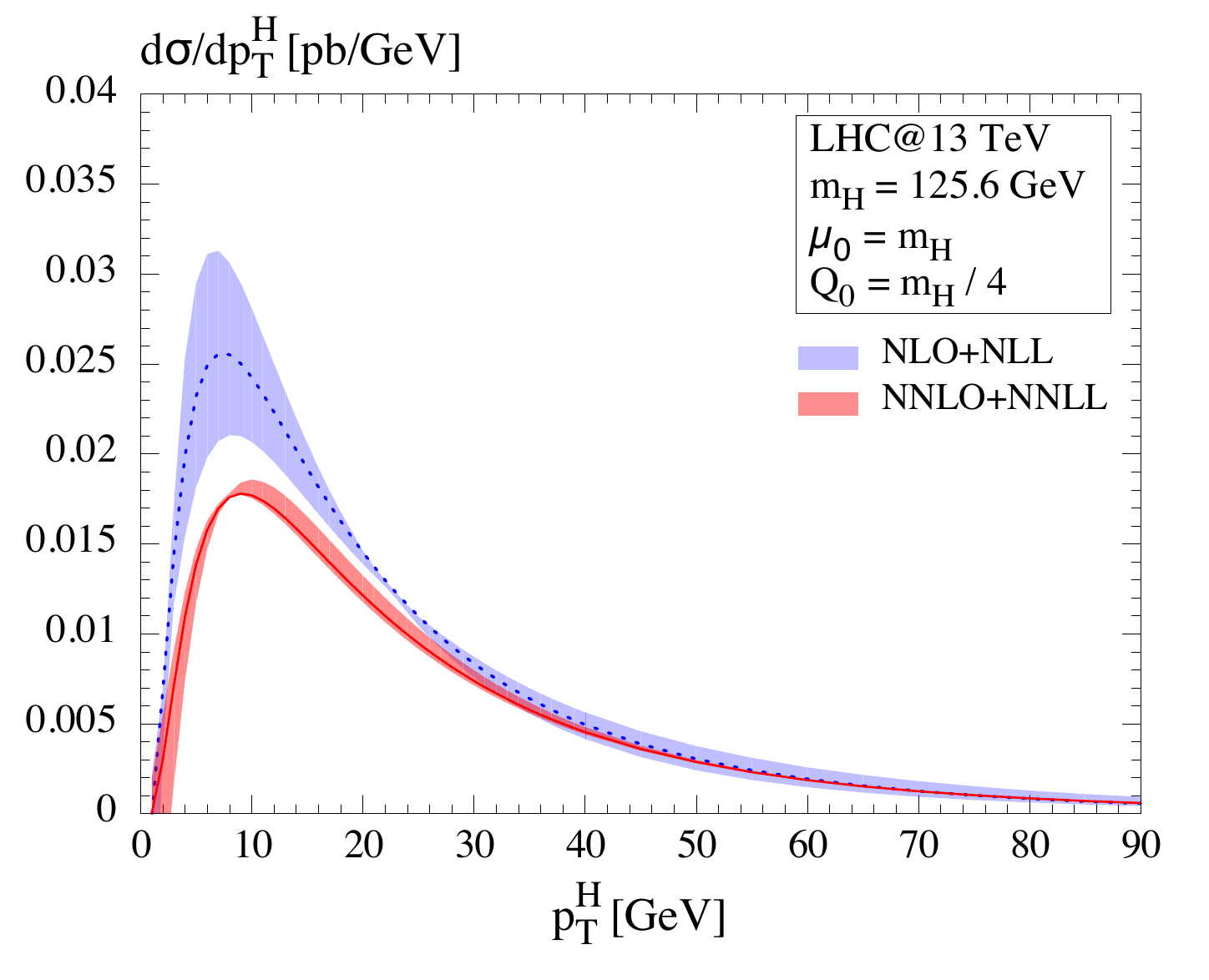}
    \parbox{.9\textwidth}{%
      \caption[]{\label{fig:Qres13}{Resummed-matched $p_T$-distribution at \nlo{}\plus\nll{} (blue, dashed line) and \nnlo{}\plus\nnll{} (red, solid line); lines: central scale choices; bands: uncertainty due to $\Qres{}$-variation. (Same as \fig{fig:Qres}, but for $\sqrt{s}=13$\,TeV.)
        }
      }}
\end{center}
\end{figure}
\begin{figure}
\begin{center}
\includegraphics[height=.385\textheight]{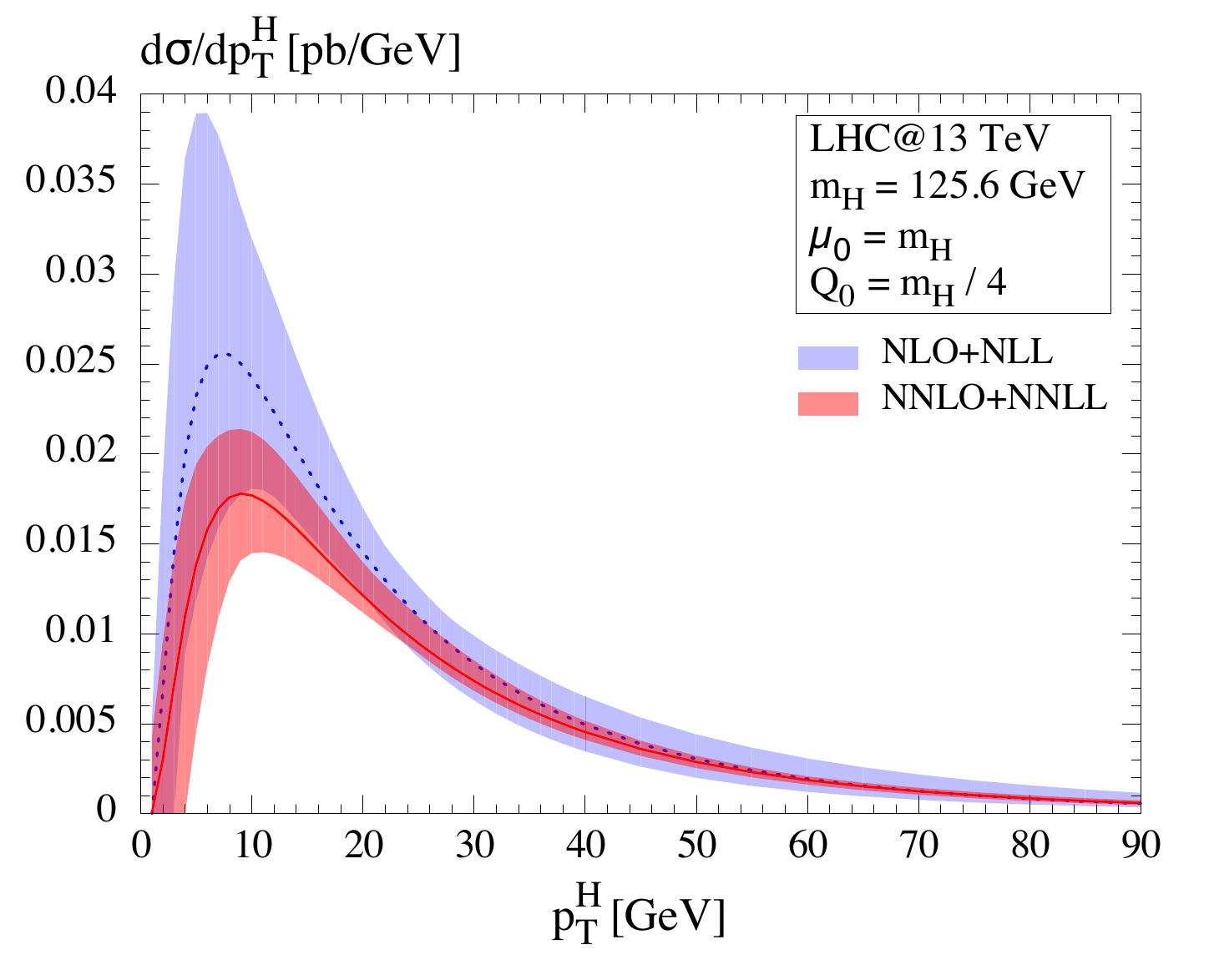}
    \parbox{.9\textwidth}{%
      \caption[]{\label{fig:all13}{\sloppy Resummed-matched $p_T$-distribution at
          \nlo{}\plus\nll{} (blue, dashed line) and \nnlo{}\plus\nnll{} (red, solid line); lines:
          central scale choices; bands: uncertainty due to variation of
          all scales. (Same as \fig{fig:all}, but for $\sqrt{s}=13$\,TeV.)}
                }}
\end{center}
\end{figure}

\newpage

\section{Results for \bld{Q_0=m_H/2}}\label{app:Q05}

In this appendix, we present the main results of the paper for a central
resummation scale of $Q_0=M/2$.

\begin{figure}[h]
\begin{center}
    \begin{tabular}{cc}
     \hspace{-0.45cm}
     \mbox{\includegraphics[height=.3\textheight]{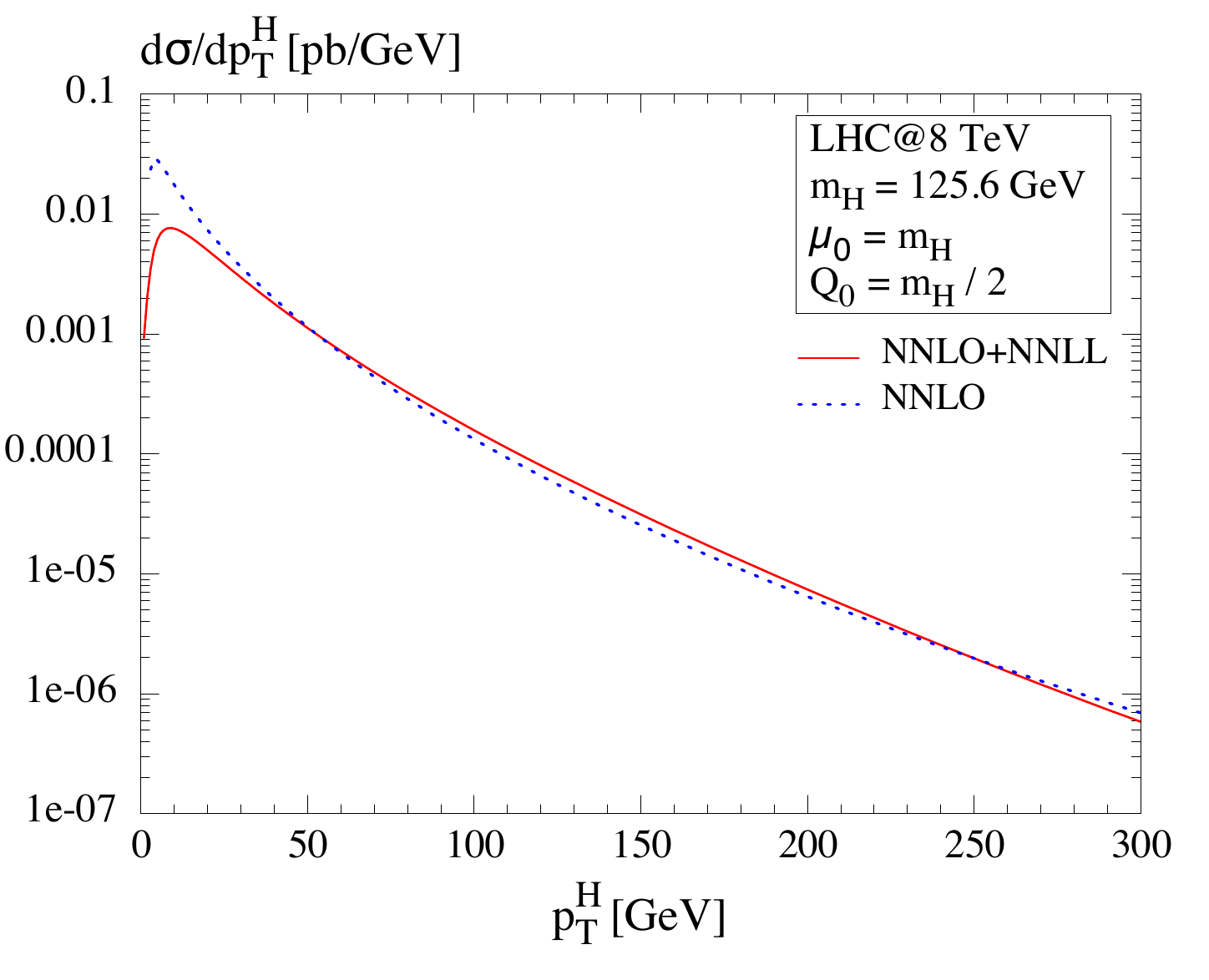}} &
     \mbox{\includegraphics[height=.3\textheight]{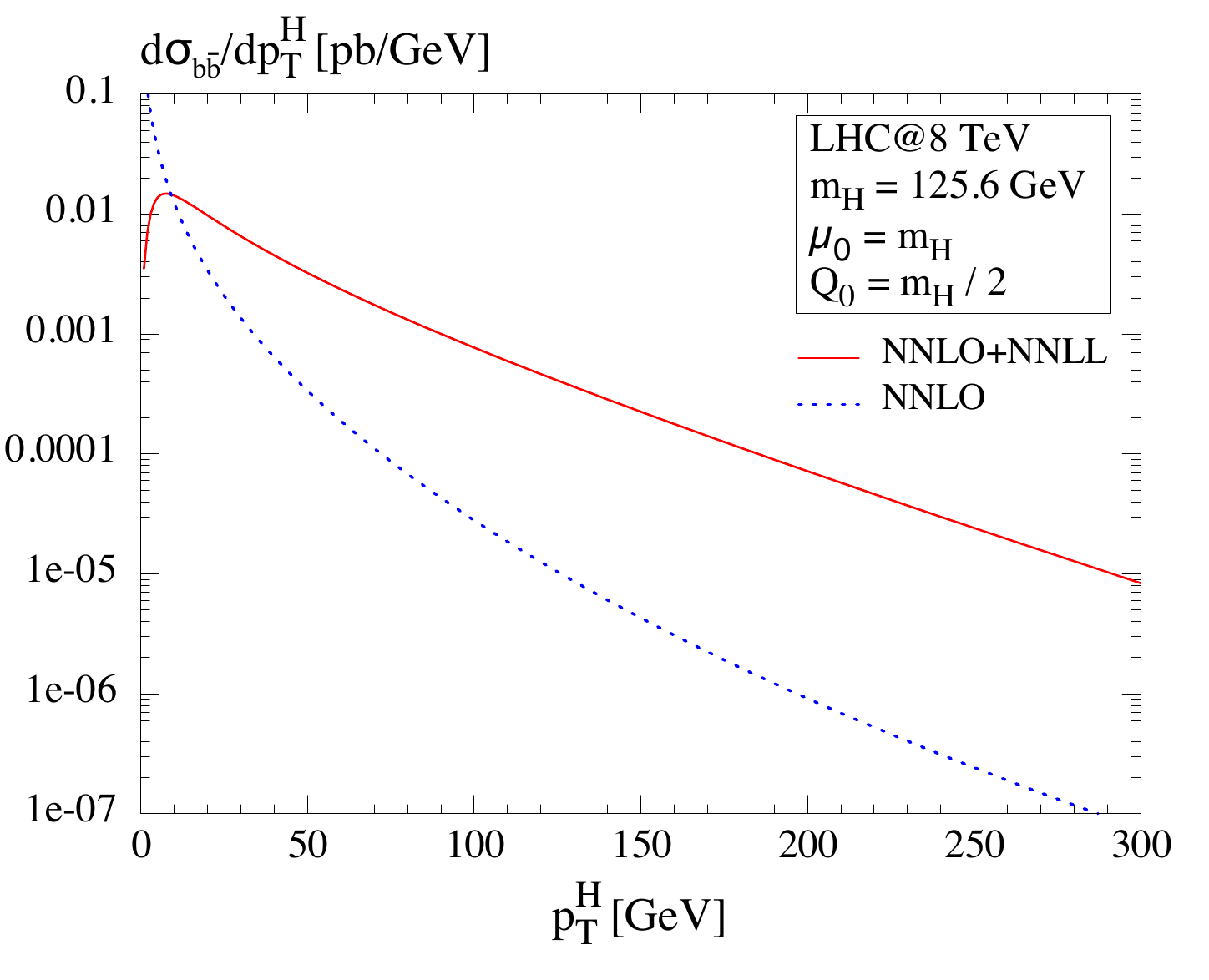}}
            \\
     \hspace{0.15cm} (a) & (b)
    \end{tabular}
    \parbox{.9\textwidth}{%
      \caption[]{\label{fig:highpTQ05}{(a) Transverse momentum spectrum at
          \nnlo{} (blue, dashed line) and at \nnlo\plus\nnll{} (red, solid line)
          for the central scales; (b) only the $b\bar{b}$ channel for
          that quantity. (Same as \fig{fig:highpT}, but for $Q_0=M/2$.) }}}
\end{center}
\end{figure}

\begin{figure}
\begin{center}
\includegraphics[height=.385\textheight]{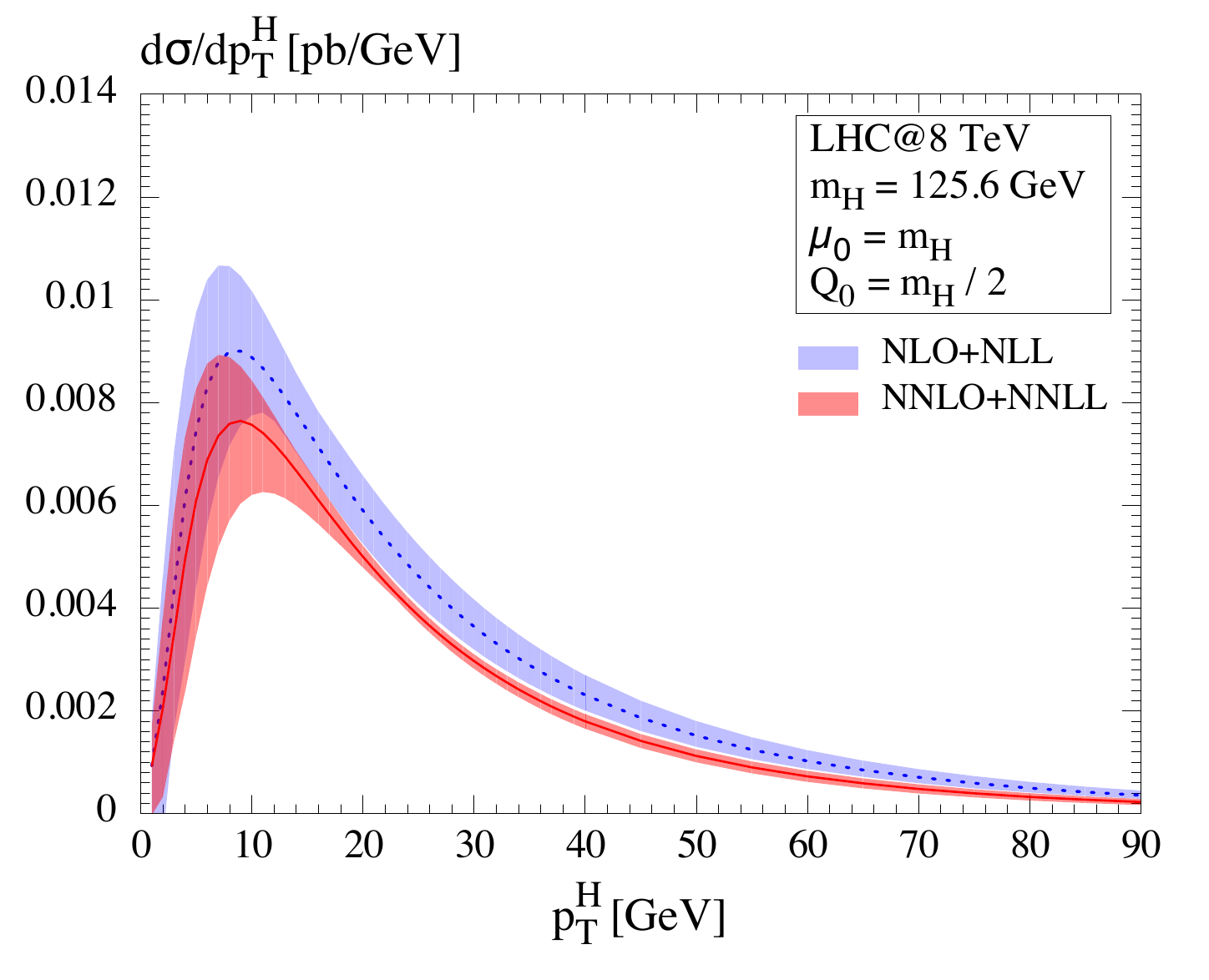}
    \parbox{.9\textwidth}{%
      \caption[]{\label{fig:muFmuRQ05}{Resummed-matched $p_T$-distribution at
          \nlo{}\plus\nll{} (blue, dashed) and \nnlo{}\plus\nnll{} (red, solid); lines:
          central scale choices; bands: uncertainty due to
          $\muF{},\muR{}$-variation.  (Same as \fig{fig:muFmuR}, but
          for $Q_0=M/2$.) } }}
\end{center}
\end{figure}

\begin{figure}
\begin{center}
\includegraphics[height=.385\textheight]{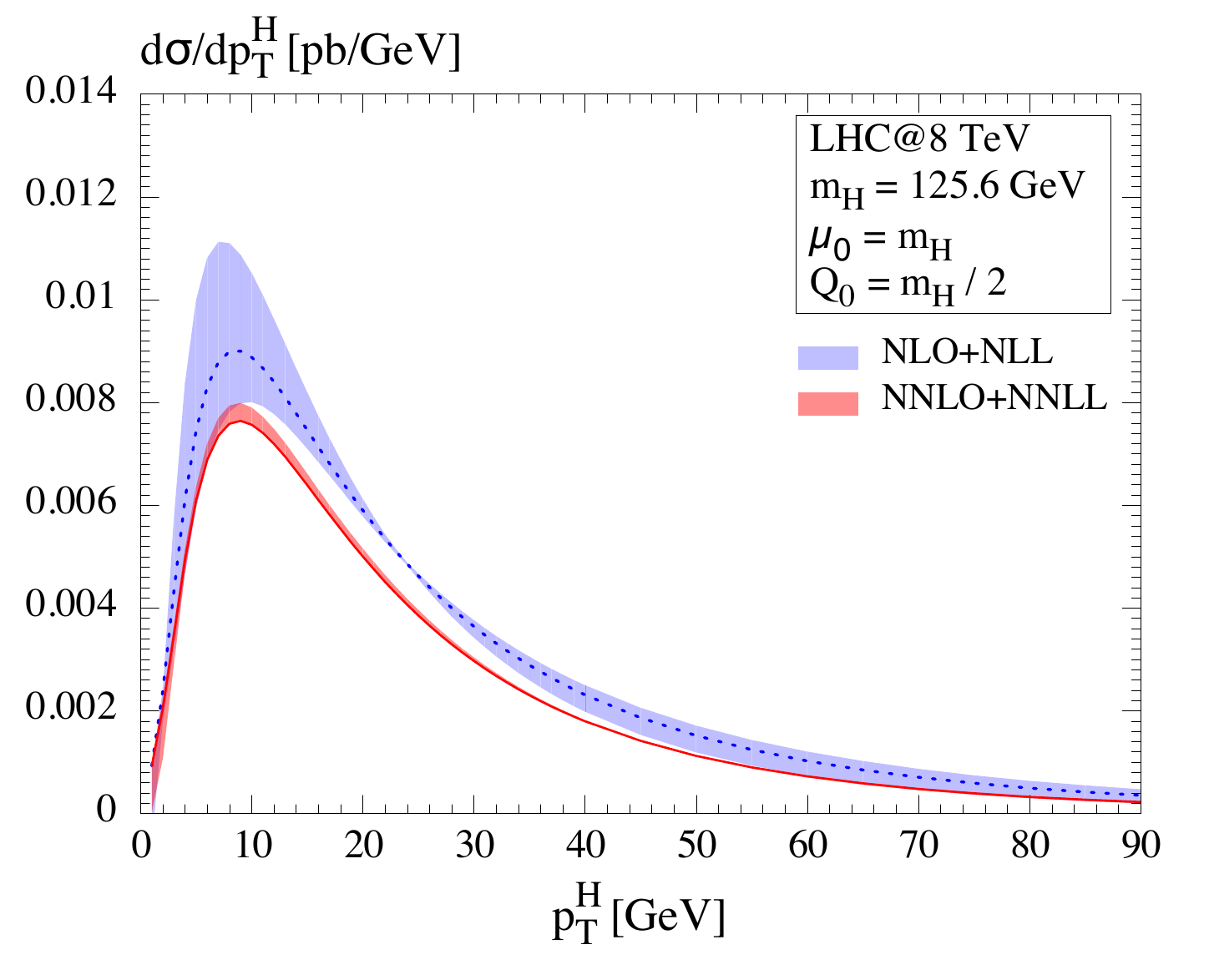}
    \parbox{.9\textwidth}{%
      \caption[]{\label{fig:QresQ05}{Resummed-matched $p_T$-distribution at
          \nlo{}\plus\nll{} (blue, dashed line) and \nnlo{}\plus\nnll{} (red, solid line); lines:
          central scale choices; bands: uncertainty due to
          $\Qres{}$-variation.  (Same as \fig{fig:Qres}, but for
          $Q_0=M/2$.) } }}
\end{center}
\end{figure}

\begin{figure}
\begin{center}
\includegraphics[height=.385\textheight]{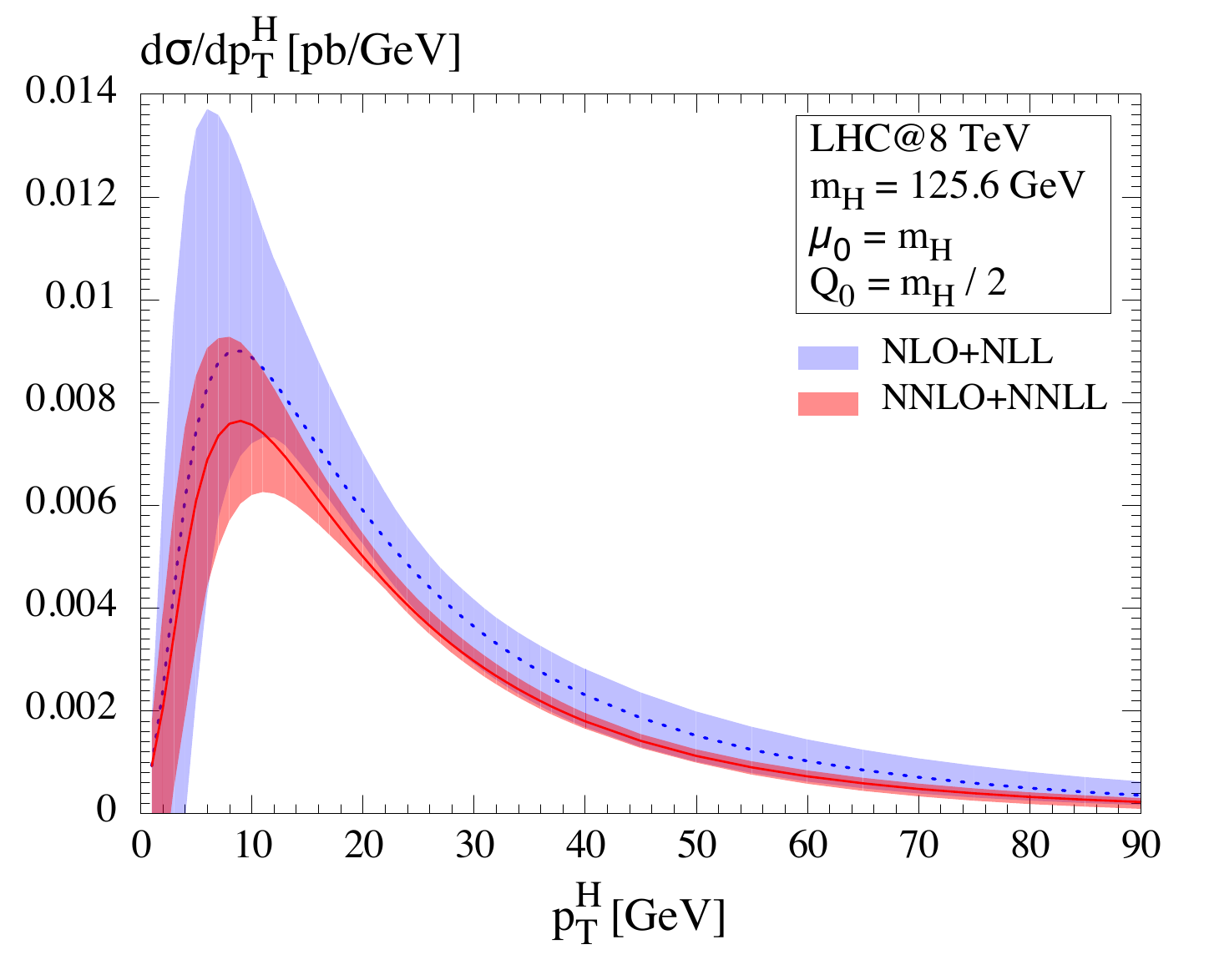}
    \parbox{.9\textwidth}{%
      \caption[]{\label{fig:allQ05}{\sloppy Resummed-matched $p_T$-distribution at
          \nlo{}\plus\nll{} (blue, dashed line) and \nnlo{}\plus\nnll{} (red, solid line); lines:
          central scale choices; bands: uncertainty due to variation of
          all scales.  (Same as \fig{fig:all}, but for $Q_0=M/2$.) } }}
\end{center}
\end{figure}

\begin{figure}
\begin{center}
\includegraphics[height=.385\textheight]{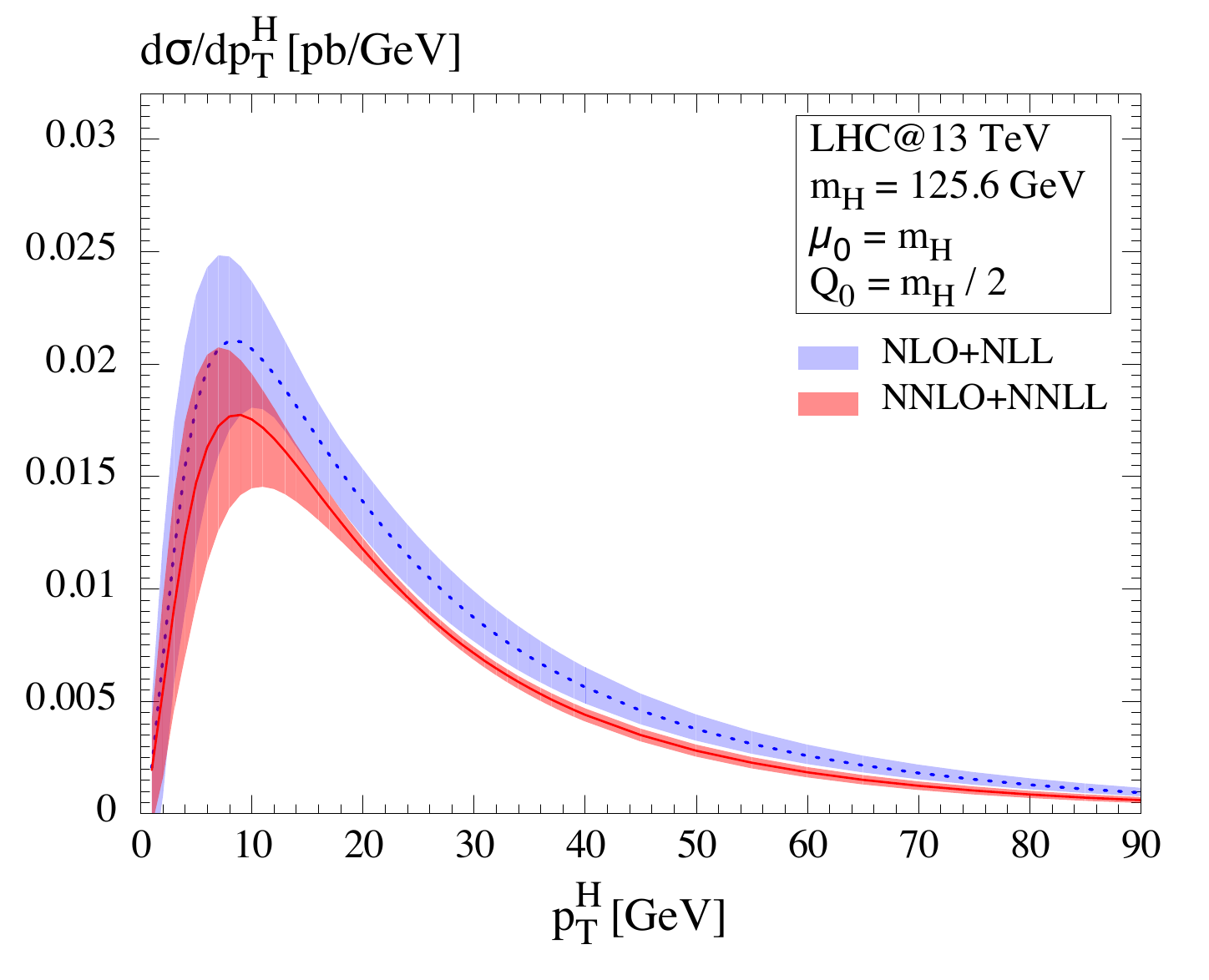}
    \parbox{.9\textwidth}{%
      \caption[]{\label{fig:muFmuR13Q05}{Resummed-matched $p_T$-distribution at
          \nlo{}\plus\nll{} (blue, dashed line) and \nnlo{}\plus\nnll{} (red, solid line); lines:
          central scale choices; bands: uncertainty due to
          $\muF{},\muR{}$-variation.  (Same as \fig{fig:muFmuR13}, but
          for $Q_0=M/2$.)  } }}
\end{center}
\end{figure}

\begin{figure}
\begin{center}
\includegraphics[height=.385\textheight]{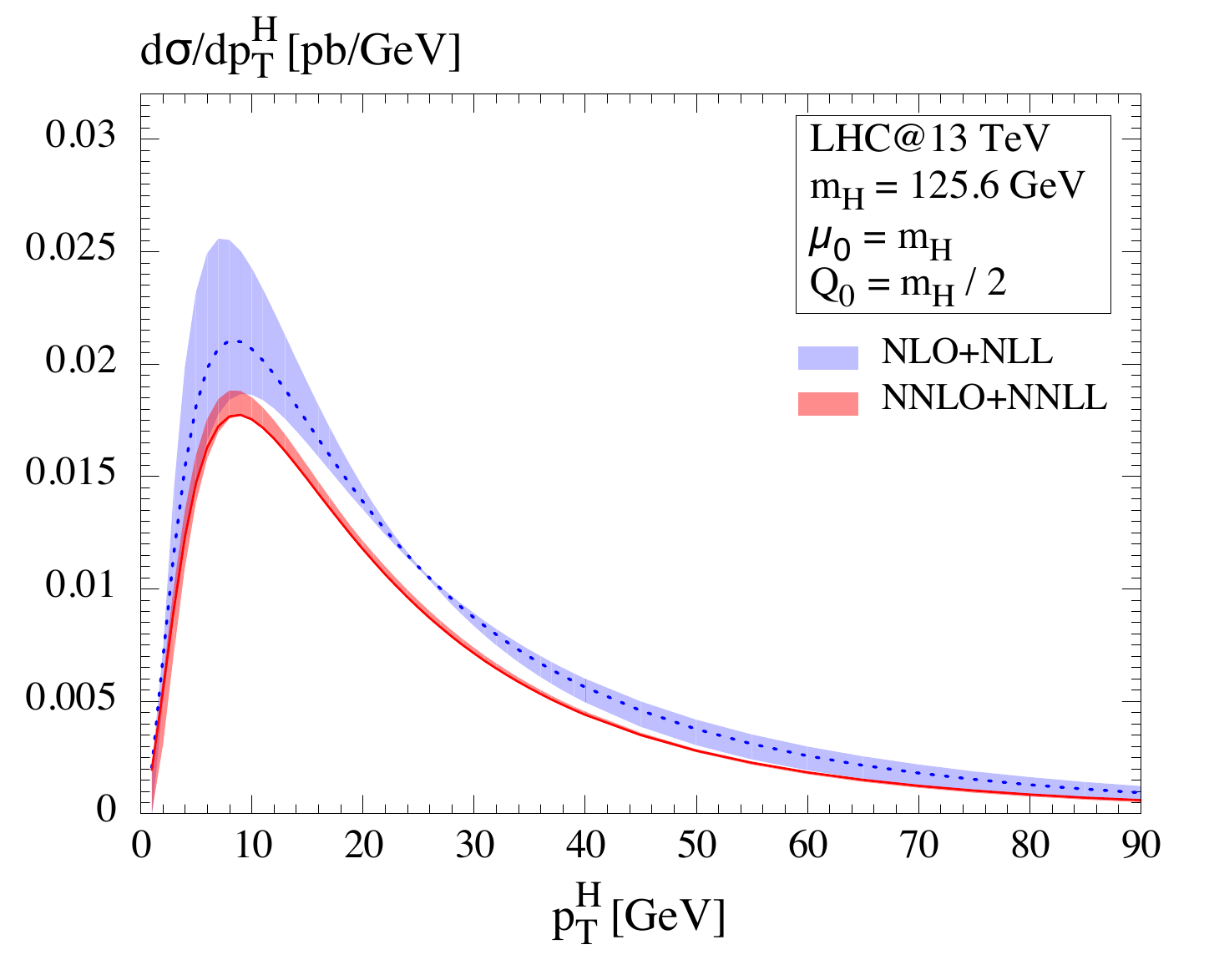}
    \parbox{.9\textwidth}{%
      \caption[]{\label{fig:Qres13Q05}{
        Resummed-matched $p_T$-distribution at \nlo{}\plus\nll{} (blue, dashed line) and \nnlo{}\plus\nnll{} (red, solid line); lines: central scale choices; bands: uncertainty due to $\Qres{}$-variation. (Same as \fig{fig:Qres13}, but for $Q_0=M/2$.)
        }
      }}
\end{center}
\end{figure}

\begin{figure}
\begin{center}
\includegraphics[height=.385\textheight]{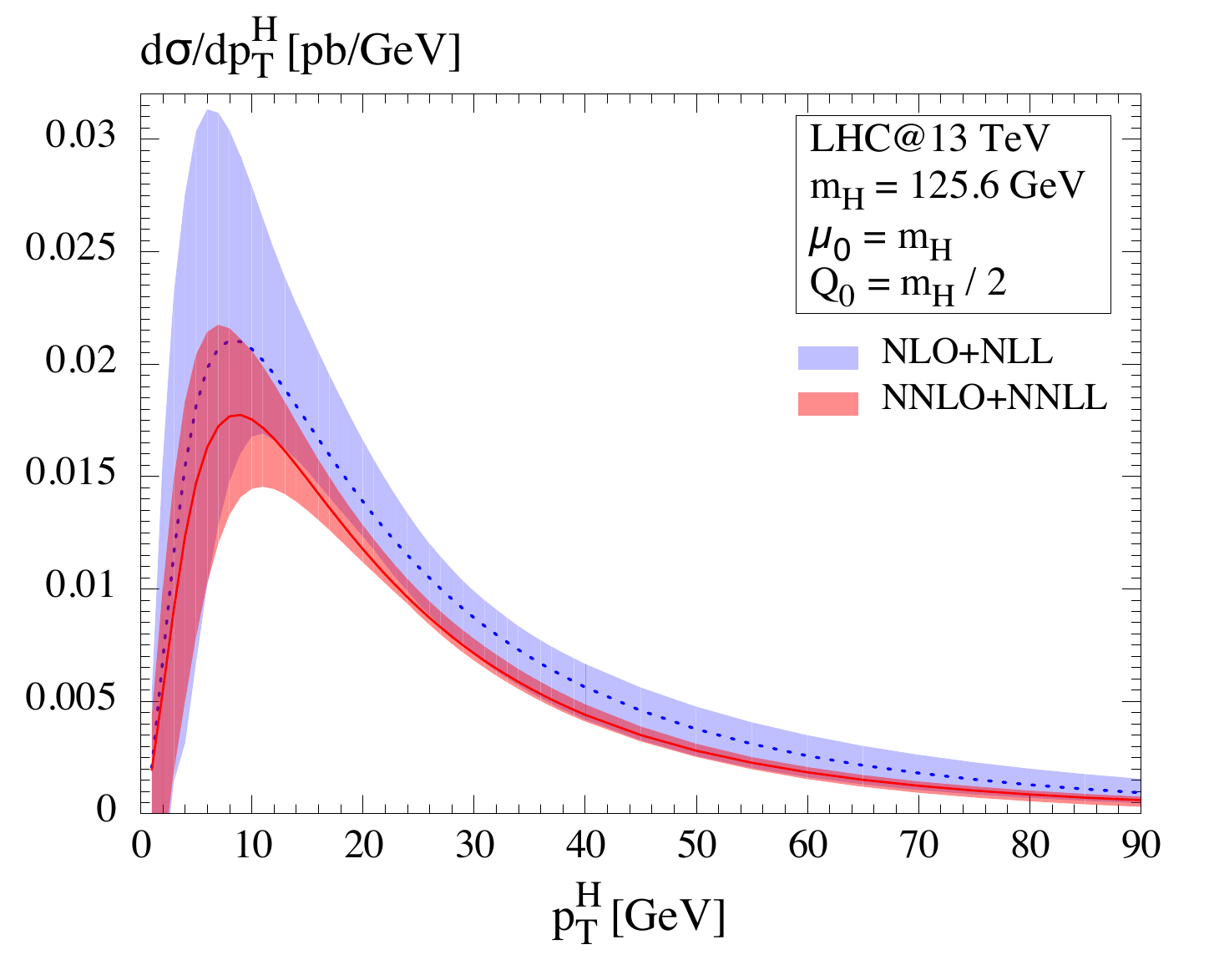}
    \parbox{.9\textwidth}{%
      \caption[]{\label{fig:all13Q05}{\sloppy Resummed-matched $p_T$-distribution at
          \nlo{}\plus\nll{} (blue, dashed line) and \nnlo{}\plus\nnll{} (red, solid line); lines:
          central scale choices; bands: uncertainty due to variation of
          all scales.  (Same as \fig{fig:all13}, but for $Q_0=M/2$.)}
                }}
\end{center}
\end{figure}

\pagebreak
\def\app#1#2#3{{\it Act.~Phys.~Pol.~}\jref{\bf B #1}{#2}{#3}}
\def\apa#1#2#3{{\it Act.~Phys.~Austr.~}\jref{\bf#1}{#2}{#3}}
\def\annphys#1#2#3{{\it Ann.~Phys.~}\jref{\bf #1}{#2}{#3}}
\def\cmp#1#2#3{{\it Comm.~Math.~Phys.~}\jref{\bf #1}{#2}{#3}}
\def\cpc#1#2#3{{\it Comp.~Phys.~Commun.~}\jref{\bf #1}{#2}{#3}}
\def\epjc#1#2#3{{\it Eur.\ Phys.\ J.\ }\jref{\bf C #1}{#2}{#3}}
\def\fortp#1#2#3{{\it Fortschr.~Phys.~}\jref{\bf#1}{#2}{#3}}
\def\ijmpc#1#2#3{{\it Int.~J.~Mod.~Phys.~}\jref{\bf C #1}{#2}{#3}}
\def\ijmpa#1#2#3{{\it Int.~J.~Mod.~Phys.~}\jref{\bf A #1}{#2}{#3}}
\def\jcp#1#2#3{{\it J.~Comp.~Phys.~}\jref{\bf #1}{#2}{#3}}
\def\jetp#1#2#3{{\it JETP~Lett.~}\jref{\bf #1}{#2}{#3}}
\def\jphysg#1#2#3{{\small\it J.~Phys.~G~}\jref{\bf #1}{#2}{#3}}
\def\jhep#1#2#3{{\small\it JHEP~}\jref{\bf #1}{#2}{#3}}
\def\mpla#1#2#3{{\it Mod.~Phys.~Lett.~}\jref{\bf A #1}{#2}{#3}}
\def\nima#1#2#3{{\it Nucl.~Inst.~Meth.~}\jref{\bf A #1}{#2}{#3}}
\def\npb#1#2#3{{\it Nucl.~Phys.~}\jref{\bf B #1}{#2}{#3}}
\def\nca#1#2#3{{\it Nuovo~Cim.~}\jref{\bf #1A}{#2}{#3}}
\def\plb#1#2#3{{\it Phys.~Lett.~}\jref{\bf B #1}{#2}{#3}}
\def\prc#1#2#3{{\it Phys.~Reports }\jref{\bf #1}{#2}{#3}}
\def\prd#1#2#3{{\it Phys.~Rev.~}\jref{\bf D #1}{#2}{#3}}
\def\pR#1#2#3{{\it Phys.~Rev.~}\jref{\bf #1}{#2}{#3}}
\def\prl#1#2#3{{\it Phys.~Rev.~Lett.~}\jref{\bf #1}{#2}{#3}}
\def\pr#1#2#3{{\it Phys.~Reports }\jref{\bf #1}{#2}{#3}}
\def\ptp#1#2#3{{\it Prog.~Theor.~Phys.~}\jref{\bf #1}{#2}{#3}}
\def\ppnp#1#2#3{{\it Prog.~Part.~Nucl.~Phys.~}\jref{\bf #1}{#2}{#3}}
\def\rmp#1#2#3{{\it Rev.~Mod.~Phys.~}\jref{\bf #1}{#2}{#3}}
\def\sovnp#1#2#3{{\it Sov.~J.~Nucl.~Phys.~}\jref{\bf #1}{#2}{#3}}
\def\sovus#1#2#3{{\it Sov.~Phys.~Usp.~}\jref{\bf #1}{#2}{#3}}
\def\tmf#1#2#3{{\it Teor.~Mat.~Fiz.~}\jref{\bf #1}{#2}{#3}}
\def\tmp#1#2#3{{\it Theor.~Math.~Phys.~}\jref{\bf #1}{#2}{#3}}
\def\yadfiz#1#2#3{{\it Yad.~Fiz.~}\jref{\bf #1}{#2}{#3}}
\def\zpc#1#2#3{{\it Z.~Phys.~}\jref{\bf C #1}{#2}{#3}}
\def\ibid#1#2#3{{ibid.~}\jref{\bf #1}{#2}{#3}}
\def\otherjournal#1#2#3#4{{\it #1}\jref{\bf #2}{#3}{#4}}
\newcommand{\jref}[3]{{\bf #1}, #3 (#2)}
\newcommand{\hepph}[1]{{\tt [hep-ph/#1]}}
\newcommand{\mathph}[1]{{\tt [math-ph/#1]}}
\newcommand{\arxiv}[2]{{\tt arXiv:#1}}
\newcommand{\bibentry}[4]{#1, {\it #2}, #3\ifthenelse{\equal{#4}{}}{}{, }#4.}

\end{document}